\documentclass{article}
\usepackage{epsfig}

\begin{document}

\def\be{\begin{equation}}
\def\ee{\end{equation}}
\def\ba{\begin{eqnarray}}
\def\ea{\end{eqnarray}}

\title{Two Interacting Electrons in a Disorder Potential: Localization
  Properties\\}

\author{Jean Richert\\
Laboratoire de Physique Th\'eorique${^1}$, \\
Universit\'e Louis Pasteur, Strasbourg, France \\
and \\
Hans A. Weidenm\"uller\\
Max-Planck-Institut f\"ur Kernphysik,
D-69029 Heidelberg, \\
Germany}

\date{\today}

\maketitle

\begin{abstract}Two electrons move in a quasi one--dimensional wire
  under the influence of a short--range interaction. We restrict
  Hilbert space to those states where the two electrons are close to
  each other. Using supersymmetry, we present a complete analytical
  solution to this problem. The two--body interaction affects the
  density of states and, thereby, the localization length. We derive a
  criterion for the onset of changes of the localization length due to
  the two--body interaction.

PACS numbers: 72.15Rn 71.30+h
\end{abstract}

\footnote{$^1$}{UMR 7085, ULP/CNRS}

\newpage

\section{Introduction}
\label{int}

The eigenfunctions of electrons moving in a random disorder potential in
one or two dimensions display localization. The influence of the Coulomb
interaction (or of any other two--body interaction) on localization
properties has been an open problem for a long time. In 1994,
Shepelyansky~\cite{dima} investigated the motion of two interacting
electrons in a disordered one--dimensional wire numerically. He found
that the two--body interaction may strongly increase the localization
length, independently of the sign of the interaction. This surprising
result was rederived by Imry~\cite{imry} who used Thouless' block
scaling picture. Further numerical work~\cite{frahm1,oppen,kim,opp,
  rom,frahm2,song} in one dimension aimed at a better understanding
and quantification of these findings, and at settling a controversy
which had arisen. As a result, the effect is now well established. In
addition, Song and von Oppen~\cite{song} have demonstrated that it
depends strongly on the distance between the two electrons and is
strongest when these remain close together.

It is obviously desirable to extend these investigations to
two--dimensional systems. At present, numerical approaches do not seem
capable of handling the ensuing difficulties. However, the work of
Refs.~\cite{dima,imry,frahm2} yields an analytical expression for the
change of the localization length $\zeta(U)$ under the influence of a
two--body interaction with typical matrix element $U$. The result
\be
\frac{\zeta(U)}{\zeta(0)} - 1 \propto \biggl( \frac{U}{B} \biggr)^2 \ ,
\label{0}
\ee
with $B$ the bandwidth due to the disorder potential, is independent
of the sign of the two--body interaction and is claimed to hold both
in one and in two dimensions. Numerical results in one
dimension~\cite{opp}, although in line with the sign--independence,
do not fully support the $U^2$ dependence predicted analytically.
Moreover, the effect is numerically found to be strongest when the two
electrons move at short distance~\cite{song}, a feature which is also
not covered by the analytical expression. Finally, the
result~(\ref{0}) has either been derived heuristically~\cite{imry}, or
it is based on statistical assumptions~\cite{frahm2} which, although
plausible, are not obviously valid.

This situation calls for a novel approach to the problem.
In the present paper, we extend the study of localization properties
of two interacting electrons to the case where the electrons move in a
quasi one--dimensional disordered wire. This case is obviously more
realistic (and numerically much more difficult) than that of the
strictly one--dimensional problem. Therefore, we aim at an
analytical treatment. The electrons interact via a short--range
interaction. The realistic case of the Coulomb interaction is supposed
to be included in our treatment because screening removes the
long--range part of that interaction. We use an analytical method
developed recently for the study of the $k$--body embedded ensembles
of random matrices~\cite{ben}. Motivated by the numerical results of
Ref.~\cite{song}, we reduce Hilbert space and keep only states where
the two electrons stay reasonably close together. This leads to a
non--linear sigma model. We determine the saddle--point manifold and
thereby the influence of the two--body interaction on localization. We
investigate all terms in the loop expansion up to second order in the
hopping matrix elements and show that these provide the expected
renormalization of the saddle--point values. We derive a criterion for
the onset of interaction--induced changes in the localization length.

A brief account of this work omitting most technical details was
published in Ref.~\cite{rich}.

\section{Model}
\label{mod}

We treat the quasi one--dimensional wire in the manner introduced in
Ref.~\cite{iid}. The wire of length $L$ is considered as being divided
into $K$ slices labelled $a,b,c,\ldots,K$. The surface connecting
neighboring slices is transverse to the direction of the current
through the wire. We eventually consider the limit in which the
longitudinal length of the slices tends to zero and, simultaneously,
$K$ tends to infinity. In each slice the actual disorder is replaced
by a random single--particle Hamiltonian. To simplify the notation and
algebra, we consider the case of unitary symmetry. Hopping matrix
elements connect neighboring slices and allow an electron to move from
slice $a$ to slices $a \pm 1$. This model has been successfully used to
calculate universal conductance fluctuations~\cite{iid} and, later, the 
localization properties of both, the average conductance and the
variance of the conductance~\cite{zir,mir}. The interaction between
the two electrons moving in the disordered wire vanishes unless both
electrons occupy the same slice.

In slice $a$ we take an arbitrary basis of single--particle states
labelled $| a j \rangle$, with $j = 1,2,\ldots,l$. We later take the limit
$l \rightarrow \infty$. The associated creation and annihilation operators
are denoted by $\alpha^{\dagger}_{aj}$ and $\alpha_{aj}$, respectively.
The Hilbert space of the two--electron problem is spanned by the
orthonormal states $|ai bj \rangle$ defined by
\be
|ai bj \rangle = \alpha^{\dagger}_{a i} \alpha^{\dagger}_{b j} | 0
\rangle \ ,
\label{eq0}
\ee
where $| 0 \rangle$ denotes the vacuum. This definition is unique if we
require $a \leq b$ and, for $a = b$, $i < j$. For $a < b$ fixed, the
number $N_{ab}$ of states is $l^2$ while for $a = b$, we have $N_{aa} =
l(l-1)/2$. We always take $l \gg 1$ and use $N_{aa} = l^2/2$.

The Hamiltonian $H$ is the sum of three terms,
\be
H = H_0 + H_1 + H_2 \ .
\label{eq00}
\ee
The single--particle Hamiltonian
\be
H_0 = \sum_{aij} h^{(a)}_{ij} \alpha^{\dagger}_{ai} \alpha_{aj}
\label{eq1}
\ee
describes the motion of the electron within each slice under the influence
of the disorder potential. The Hermitean matrices $h^{(a)}$ are members of
the Gaussian unitary ensemble of random matrices. They have dimension $l$,
mean value zero and second moments given by
\be
\overline{h^{(a)}_{ij} h^{(a')*}_{i'j'}} = \frac{\lambda^2}{l}
\delta_{a a'} \delta_{i i'} \delta_{j j'} \ .
\label{eq2}
\ee
The overbar denotes the ensemble average. The parameter $\lambda$ has the
dimension of an energy and is independent of the slice label $a$. The
single--particle spectrum in each slice has the shape of a semicircle
with radius $2 \lambda$. Hopping between neighboring slices is described
by the Hamiltonian
\be
H_{1} = \sum_{a ij} [v^{(a)}_{ij} \alpha^{\dagger}_{ai} \alpha_{a+1 j}
+ h.c.] \ .
\label{eq3}
\ee
It was shown in Ref.~\cite{iid} that the hopping matrix elements
$v^{(a)}$ may either be chosen as Gaussian--distributed complex random
variables with mean value zero and a second moment
$\overline{v^{(a)}_{ij} v^{(a')*}_{i'j'}} = \frac{v^2}{l} \delta_{a
  a'} \delta_{i i'} \delta_{j j'}$ or, equivalently, as elements of a
constant diagonal matrix,
\be
v^{(a)}_{ij} = v \ \delta_{i j} \ . 
\label{eq4}
\ee
Here, $v$ is real and independent of the slice label $a$. In the
present context, we adopt the second alternative. The two--body
interaction $H_{2}$ is fixed (not random) and has the same matrix
elements within each slice,
\be
H_2 = \sum_a \sum_{i<j,i'<j'} w_{iji'j'} \alpha^{\dagger}_{ai}
\alpha^{\dagger}_{aj} \alpha_{aj'} \alpha_{ai'} \ .
\label{eq5}
\ee
The two--body matrix elements are antisymmetric in the pairs $(ij)$
and $(i'j')$ and Hermitean.

We wish to determine how the two--body interaction~(\ref{eq5}) affects
localization properties. The effect (if any) will be most dramatic if
both electrons stay close to each other as they traverse the wire. In
order to focus on this point, and in order to simplify the calculation,
we reduce the Hilbert space of two--particle states. We admit only
states where both electrons are in the same slice, or in
neighboring slices. Thus, of all the states introduced in
Eq.~(\ref{eq0}), we keep only the ones with $b = a$ and with $b = (a +
1)$. Without this omission, our model would be very general. It is,
therefore, important to address the physical significance of this
simplification. The numerical work of Ref.~\cite{song} in one dimension
has clearly demonstrated that the change of the localization length
caused by the two--body interaction is biggest when the two electrons
which move through the one--dimensional wire, keep the shortest distance
from each other. This is our motivation for the constraint artificially
imposed in our model. It might have been desirable to loosen the
constraint and to allow the electrons to keep a maximum distance given
by the localization length in the absence of disorder. However, this
was technically impossible. Therefore, we cannot claim that our
results are quantitatively correct. However, in view of the work of
Ref.~\cite{song}, we believe that we obtain qualitatively correct
answers. The shortcoming of our approach -- the reduction of Hilbert
space -- must be weighed against the fact that with this
simplification, we are able to obtain a complete analytical solution
of and a physically transparent answer to the problem.

We aim at a comparison of the localization properties of this simplified
system when the two--body interaction~(\ref{eq5}) is either turned off
or fully present. Localization properties can be read off the two--point
correlation function
\be
C(n) = \overline{|\langle a (a+1) \mu_0| (E^+ - H)^{-1} | (a+n) (a+n+1)
\nu_0 \rangle|^2} \ .
\label{a1}
\ee
The indices $\mu_0$ and $\nu_0$ are defined in Eq.~(\ref{eq6}) below.
To be independent of edge effects, we choose $1 \ll a \ll (a+n) \ll K$.
For large $n$, $C(n)$ should decay exponentially with $n$. In proper
units, the coefficient in the exponent defines the localization length.
We calculate $C(n)$ using supersymmetry~\cite{efe,ver}.

\section{Second Moment of the One--Body \\ Hamiltonian
  $H_0$. Eigenvalue Expansion}
\label{mom}

The matrix elements of the single--particle Hamiltonian $H_0$ in the
Hilbert space of two--electron states~(\ref{eq0}) with $b=a$ and $b =
(a + 1)$ are Gaussian--distributed random variables with mean value
zero. Therefore, the distribution of these matrix elements is fully
determined by their second moments. We follow Refs.~\cite{ben} and
consider these second moments as matrices in the product space of two
Hilbert--space vectors. We use this representation to derive the
eigenvalue expansion of the second moments. This will allow us to
apply the supersymmetry technique.

The operator $H_0$ does not change the number of particles per slice.
Therefore, the operator $\alpha^{\dagger}_{a k} \alpha_{a k'}$ occurs
in the following three types of matrix elements: $\langle (a-1) i a j
| H_0 | (a-1) i' a j' \rangle$,  $\langle a i a j | H_0 | a i' a j'
\rangle$, and $\langle a i (a+1) j | H_0 | a i' (a+1) j' \rangle$,
with $a = 1, \ldots, K$. Here and in the sequel, we write the adjoint
of the Hilbert vector $| a i b j \rangle$ as $ \langle a i b j |$. We
simplify the notation by grouping the state labels into pairs,
\be
\mu = (i_1 j_1), \ \sigma = (i'_1 j'_1), \ \rho = (i_2 j_2), \ \nu =
(i'_2 j'_2) \ .
\label{eq6}
\ee
Thus, the Hilbert vector $| a i b j \rangle$ is written as $|a b \mu
\rangle$. In the sequel, we refer to the set of states $|a b \mu
\rangle$ encompassing all values of $\mu$ as to a box.

There are nine types of terms contributing to the second moment.
Three of them have ``diagonal form'' and are given by
\ba
&&\frac{l}{\lambda^2} \ \overline{ \langle (a-1) a \mu | H_0 | (a-1) a
\sigma \rangle \langle (a-1) a \rho | H_0 | (a-1) a \nu \rangle }
\nonumber \\
&&\qquad = {\cal A}^{(-1)}_{\mu \nu ; \rho \sigma} 
= \sum_{c k k'} \langle (a-1) a \mu | \alpha^{\dagger}_{c k} \alpha_{c
k'} | (a-1) a \sigma \rangle \nonumber \\
&&\qquad \qquad \qquad \qquad \times \langle (a-1) a \rho |
\alpha^{\dagger}_{c k'} \alpha_{c k} | (a-1) a \nu \rangle \ ,
\label{eq7a}
\ea
\ba
&&\frac{l}{\lambda^2} \ \overline{ \langle a a \mu | H_0 | a a \sigma
\rangle \langle a a \rho | H_0 | a a \nu \rangle } =
{\cal A}^{(0)}_{\mu \nu ; \rho \sigma} \nonumber \\
&&\qquad = \sum_{k k'} \langle a a \mu | \alpha^{\dagger}_{a k}
\alpha_{a k'} | a a \sigma \rangle \langle a a \rho |
\alpha^{\dagger}_{a k'} \alpha_{a k} | a a \nu \rangle \ ,
\label{eq7b}
\ea
\ba
&&\frac{l}{\lambda^2} \ \overline{ \langle a (a+1) \mu | H_0 | a (a+1)
\sigma \rangle \langle a (a+1) \rho | H_0 | a (a+1) \nu \rangle }
\nonumber \\
&&\qquad = {\cal }{\cal A}^{(+1)}_{\mu \nu ; \rho \sigma} = \sum_{c k
  k'} \langle a (a+1) \mu | \alpha^{\dagger}_{c k} \alpha_{c k'} |
a (a+1) \sigma \rangle \nonumber \\ 
&&\qquad \qquad \qquad \qquad \times \langle a (a+1) \rho | 
\alpha^{\dagger}_{c k'} \alpha_{c k} | a (a+1) \nu \rangle \ .
\label{eq7c}
\ea
The sum over $c$ in Eqs.~(\ref{eq7a},\ref{eq7c}) runs over $((a-1),a)$
and $(a,(a+1))$, respectively. We observe that ${\cal A}^{(+1)}_{\mu
  \nu ; \rho \sigma}$ is obtained from ${\cal A}^{(-1)}_{\mu \nu ;
  \rho \sigma}$ by shifting $a \rightarrow (a+1)$. Therefore, we
consider only ${\cal A}^{(-1)}_{\mu \nu ; \rho \sigma}$ and ${\cal
  A}^{(0)}_{\mu \nu ; \rho \sigma}$ in the sequel. The
``non--diagonal'' contributions to the second moment are
\ba
&&\frac{l}{\lambda^2} \ \overline{ \langle (a-1) a \mu | H_0 | (a-1) a
\sigma \rangle \langle a a \rho | H_0 | a a \nu \rangle } \nonumber \\
&&\qquad = {\cal A}^{(-1,0)}_{\mu \nu ; \rho \sigma} = \sum_{k k'}
\langle (a-1) a \mu | \alpha^{\dagger}_{a k} \alpha_{a k'} | (a-1) a
\sigma \rangle \nonumber \\
&&\qquad \qquad \qquad \qquad \times \langle a a \rho |
\alpha^{\dagger}_{a k'} \alpha_{a k} | a a \nu \rangle \ ,
\label{eq8a}
\ea
\ba
&&\frac{l}{\lambda^2} \ \overline{ \langle (a-1) a \mu | H_0 |(a-1) a
\sigma \rangle \langle a (a+1) \rho | H_0 | a (a+1) \nu \rangle }
\nonumber \\
&&\qquad = {\cal A}^{(-1,+1)}_{\mu \nu ; \rho \sigma} = \sum_{k k'}
\langle (a-1) a \mu | \alpha^{\dagger}_{a k} \alpha_{a k'} | (a-1) a
\sigma \rangle \nonumber \\
&&\qquad \qquad \qquad \qquad \times \langle a (a+1) \rho |
\alpha^{\dagger}_{a k'} \alpha_{a k} | a (a+1) \nu \rangle \ ,
\label{eq8b}
\ea
\ba
&&\frac{l}{\lambda^2} \ \overline{ \langle a a \mu | H_0 | a a
\sigma \rangle \langle a (a+1) \rho | H_0 | a (a+1) \nu \rangle }
\nonumber \\
&&\qquad = {\cal A}^{(0,+1)}_{\mu \nu ; \rho \sigma} = \sum_{k k'}
\langle a a \mu | \alpha^{\dagger}_{a k} \alpha_{a k'} | a a \sigma
\rangle \nonumber \\
&&\qquad \qquad \qquad \qquad \times \langle a (a+1) \rho |
\alpha^{\dagger}_{a k'} \alpha_{a k} | a (a+1) \nu \rangle \ ,
\label{eq8c}
\ea
and the three Hermitean conjugate forms.

Eqs.~(\ref{eq8a}) to (\ref{eq8c}) show that the matrix elements of
$H_0$ taken in different boxes, are correlated. This is a trivial
consequence of the fact that one of the two electrons occupies the
same slice. The correlation causes our problem to differ from that
of the motion of a single elctron through the wire. We return to
this point in Sections~\ref{loop1} and \ref{disc} below.

In order to be able to apply the supersymmetry technique, we now
construct the eigenvalue expansions of the matrices ${\cal A}$
introduced in Eqs.~(\ref{eq7a}) to (\ref{eq8c}). We essentially follow
Ref.~\cite{ben}. We explicitly consider ${\cal A}^{(-1)}_{\mu \nu ;
  \rho \sigma}$ and ${\cal A}^{(-1,+1)}_{\mu \nu ; \rho \sigma}$ and
only state the results for the remaining matrices ${\cal A}$.

The matrix ${\cal A}^{(-1)}_{\mu \nu ; \rho \sigma}$ is Hermitean,
\be
({\cal A}^{(-1)}_{\sigma \rho ; \nu \mu })^{*} = {\cal A}^{(-1)}_{\mu
  \nu ; \rho \sigma} \ .
\label{eq9}
\ee
Therefore, ${\cal A}^{(-1)}$ can be expanded in terms of its
eigenvalues and eigenvectors. The eigenvalue equation reads
\be
\sum_{\rho \sigma} {\cal A}^{(-1)}_{\mu \nu ; \rho \sigma} \ C^{(s
  \tau)}_{\sigma \rho}(-1) = \Lambda^{(s)}(-1) \ C^{(s \tau)}_{\mu
  \nu}(-1) \ .
\label{eq10}
\ee
Here $s$ labels the different eigenvalues, and $\tau$ distinguishes
degenerate eigenvectors. The latter are normalized according to
\be
\sum_{\mu \nu} C^{(s \tau)}_{\mu \nu}(-1) (C^{(s' \tau')}_{\mu
\nu}(-1))^{\dagger} = N_{(a-1) a} \ \delta_{s s'} \delta_{\tau \tau'}
\label{eq11}
\ee
where $N_{(a-1) a}$ labels the dimension of the matrix space, see
Section~\ref{mod}. If the eigenvectors form a complete set, the
eigenvalue expansion of ${\cal A}^{(-1)}$ is given by
\be
{\cal A}^{(-1)}_{\mu \nu ; \rho \sigma} = \frac{1}{N_{(a-1) a}}
\sum_{s \tau} \Lambda^{(s)} C^{(s \tau)}_{\mu \nu}(-1) (C^{(s
  \tau)}_{\sigma \rho}(-1))^{\dagger} \ .
\label{eq12}
\ee
According to Ref.~\cite{ben} we expect that the sum over $s$ extends
over $s = 0, 1$ only. This is because $H_0$ is a one--body operator
and we deal with two electrons. (It was shown in Ref.~\cite{ben} that
for a general $k$--body operator and $m$ Fermions, the sum over $s$
runs from $0$ to $(m - k)$). Moreover, the eigenvectors to $s = 0$ are
expected not to be degenerate and generically to have the form
$C^{(0)}_{\mu \nu} \propto \delta_{\mu \nu} = \delta_{i_1 i'_2}
\delta_{j_1 j'_2}$ while the eigenvectors to $s = 1$ are generically
expected to be $(l^2 - 1)$--fold degenerate and to be given by $C^{(1
  \tau)}_{\mu \nu}(-1) \propto \langle (a-1) a \mu |
\alpha^{\dagger}_{c m} \alpha_{c m'} | (a-1) a \nu \rangle$, with $c =
(a-1), a$. The label $\tau$ stands for all combinations of indices
$(m, m')$ with $m \neq m'$. For $m = m'$, we have $C^{(1 \tau)}_{\mu
  \nu}(-1) \propto \langle (a-1) a \mu | (\alpha^{\dagger}_{c m}
\alpha_{c m} - (1/l) \sum_{n=1}^l \alpha^{\dagger}_{c m} \alpha_{c m})
| (a-1) a \nu \rangle$ with $c = (a-1), a$. Using this ansatz, we do
indeed solve the eigenvalue equation. The eigenvalues are
\be
\Lambda^{(0)}(-1) = 2l \ ; \ \Lambda^{(1)}(-1) = (2l -1) \ .
\label{eq13}
\ee
Eigenvalues with $s > 1$ vanish. Counting shows that the eigenvectors
form a complete set. Thus, the expansion~(\ref{eq12}) is established.

We turn to ${\cal A}^{(-1,0)}_{\mu \nu ; \rho \sigma}$ in
Eq.~(\ref{eq8a}). This matrix does not have the
property~(\ref{eq9}). Nevertheless, an expansion analogous to
Eq.~(\ref{eq12}) can be found. We first determine the right--hand and
left--hand eigenvectors. It is obvious that eigenvectors to $s = 0$ do
not exist. The right--hand eigenvectors $C^{(1 \tau)}(-1,0;r)$ to $s =
1$ are proportional to $\langle (a-1) a \mu | \alpha^{\dagger}_{(a-1)
m} \alpha_{a m'} | a a \nu \rangle$ and belong to eigenvalue
\be
\Lambda^{(1)}(-1,0) = l-1 \ .
\label{eq14}
\ee
The left--hand eigenvectors $C^{(1 \tau)}(-1,0;l)$ to the same
eigenvalue are proportional to $\langle a a \mu | \alpha^{\dagger}_{a
m} \alpha_{(a-1) m'} | (a-1) a \nu \rangle$. All eigenvalues with $s >
1$ vanish. Each pair of left-- and right--hand eigenvectors fulfills
orthogonality relations analogous to Eq.~(\ref{eq11}) and can be
normalized correspondingly. As a result, ${\cal A}^{(-1,0)}$ can be
written in the form
\be
{\cal A}^{(-1,0)}_{\mu \nu ; \rho \sigma} = \frac{1}{N(-1,0)} (l-1)
\sum_{\tau} C^{(1 \tau)}_{\mu \nu}(-1,0;l) C^{(1 \tau)}_{\sigma
\rho}(-1,0;r) \ .
\label{eq15}
\ee
The factor $N$ is the normalization factor. For the remaining five
matrices ${\cal A}$, we proceed analogously. The eigenvectors and
eigenvalues are listed in Appendix 1. In summary, all matrices ${\cal
  A}^{(i)}$ with 
\be
i = -1, 0, +1, (-1,0), (-1,+1), (0,+1), (+1,0), (+1,-1), (0,-1)
\label{eq15a}
\ee
possess the expansion
\be
{\cal A}^{(i)}_{\mu \nu ; \rho \sigma} = \frac{1}{N(i)} \sum_{s = 0}^1
\Lambda^{(s)}(i) \sum_{\tau} C^{(s \tau)}_{\mu \nu}(i;l) C^{(s
  \tau)}_{\rho \sigma}(i;r)) \ .
\label{eq16}
\ee
We observe that the left--and right--hand eigenvectors are Hermitean
adjoints of each other. In the sequel, we will need the explicit values
of the normalization factors $N(0) = l^2/2$ and $N(1) = l^2$.

\section{Supersymmetry. Generating Function}
\label{sup}

A discussion of the localization properties of the two--electron system
requires the knowledge of the average two--point function. We calculate
this function using Efetov's supersymmetry technique~\cite{efe} and
the notation of Ref.~\cite{ver}. The technique has found wide
applications in the theory of disordered solids. Therefore, we do not
give many details. We confine ourselves to indicating where the present
treatment differs from that of Refs.~\cite{ben}.

The generating functional $Z$ is an integral over commuting and
anticommuting variables. These are arranged in the form of the graded
vectors $\Psi_{a b \mu ; \alpha}$ which carry the labels $(a b)$ of
the boxes, the running label $\mu$ within each box, and the label
$\alpha$ which for every value of $(a b \mu)$ distinguishes two
complex commuting ($\alpha = 1,3$) and two anticommuting ($\alpha
= 2,4$) integration variables. The functional $Z$ contains in the
exponent of the integrand a term arising from $H_0$. This term has
the form
\be
- \frac{i}{2} \sum_{a b} \sum_{\mu \nu} \Psi^{*}_{a b \mu} L^{1/2}
\langle a b \mu | H_0 | a b \nu \rangle L^{1/2} \Psi_{a b \nu} \ .
\label{eq17}
\ee
The diagonal graded matrix $L$ is given by ${\rm diag}(1,1,-1,-1)$.
The graded indices have been
omitted in Eq.~(\ref{eq17}). After averaging over the ensemble
$\{H_0\}$, the generating functional contains in the exponent a term
given by $1/2$ times the mean value of the square of the
expression~(\ref{eq17}). This term is worked out in Appendix 2. As a
result, we find
\ba
&&\frac{1}{2} \ \overline{\biggl (- \frac{i}{2} \sum_{a b} \sum_{\mu
\nu} \Psi^{*}_{a b \mu} L^{1/2} \langle a b \mu | H_0 | a b \nu
\rangle L^{1/2} \Psi_{a b \nu} \biggr )^2} \nonumber \\
&=&\frac{1}{2 l^2} \sum_{a=1}^K {\rm trg} [ ( A^{(a a)}(0))^2] +
\frac{1}{4 l^2} \sum_{a=1}^{K-1} {\rm trg} [ ( A^{(a (a+1))}(0))^2]
\nonumber \\
&+& \frac{1}{4 l^2} \sum_{a=1}^K \sum_{\tau} {\rm trg} [ ( A^{(a
  a; \tau)}(1))^2] + \frac{1}{4 l^2} \sum_{a=1}^{K-1} \sum_{\tau} {\rm
  trg} [ ( A^{(a (a+1); \tau)}(1))^2] \nonumber \\
&+& \frac{1}{4} \sum_{a = 2}^K N(-1,0)^{-1} \sum_{\tau} {\rm trg}
[A^{((a-1) a; \tau)}(-1,0;l) A^{((a-1) a; \tau)}(-1,0;r) \nonumber \\
&+& \frac{1}{4} \sum_{a = 2}^{K-1} N(-1,1)^{-1} \sum_{\tau} {\rm trg}
[A^{((a-1) (a+1); \tau)}(-1,1;l) A^{((a-1) (a+1); \tau)}(-1,1;r)
\nonumber \\
&+& \frac{1}{4} \sum_{a = 1}^{K-1} N(1,0)^{-1} \sum_{\tau} {\rm trg}
[A^{(a (a+1); \tau)}(0,1;l) A^{(a (a+1); \tau)}(0,1;r) \ .
\label{eq17a}
\ea
The symbols $A$ appearing on the right--hand side denote graded
matrices which are defined in Appendix 2. Each of the matrices $A$ is
bilinear in the integration variables $\Psi$. As a consequence, the
expression~(\ref{eq17a}) is quartic in the $\Psi$'s. To obtain
bilinear expressions and work out the $\Psi$--integrals, one usually
performs a Hubbard--Stratonovich (HS) transformation. Here, this step
is slightly more complicated than is the case usually because some of
the terms in expression~(\ref{eq17a}) contain bilinear expressions in
(rather than squares of) the matrices $A$. To overcome this problem,
we recall that -- with proper identification of the indices $\tau$ --
the matrices $C^{1 \tau}_{\mu \nu}(-1,0;r)$ and $C^{1 \tau}_{\mu
  \nu}(-1,0;l)$ are Hermitean adjoints of each other, see
Eqs.~(\ref{a100}). Therefore, the matrices
\ba
D^{((a-1) a; \tau)} &=& \frac{1}{2} [ A^{((a-1) a; \tau)}(-1,0;r) -
A^{((a-1) a; \tau)}(-1,0;l) ] \ , \nonumber \\
E^{((a-1) a; \tau)} &=& \frac{1}{2i} [ A^{((a-1) a; \tau)}(-1,0;r) +
A^{((a-1) a; \tau)}(-1,0;l) ]
\label{eq32}
\ea
have the same symmetry properties as the matrices $A^{(aa)}(0)$. On
the other hand, using the cyclic invariance of the trace, we have
\ba
&&{\rm trg} [ A^{((a-1) a; \tau)}(-1,0;l) A^{((a-1) a; \tau)}(-1,0;r)
] \nonumber \\
&&\qquad = {\rm trg} [ (i D^{((a-1) a; \tau)})^2 + ( i E^{((a-1) a;
  \tau)})^2 ] \ .
\label{eq33}
\ea
Therefore, the usual HS transformation can now be used for $D$ and
$E$. We proceed correspondingly for $A^{((a-1) (a+1); \tau)}(-1,1;r)$
and $A^{(a (a+1); \tau)}(0,1;r)$. For every possible combination of
slice labels $(a,b)$ and for each of the graded matrices $A$ we
introduce a corresponding graded $4 \times 4$ $\sigma$--matrix with
the following correspondence:
\ba
A^{(ab)} &\leftrightarrow& \sigma^{(ab)} \ , \nonumber \\
A^{(ab; \tau)} &\leftrightarrow& \sigma^{(ab; \tau)} \ , \nonumber \\
D^{(ab; \tau)} &\leftrightarrow& \sigma^{(ab; \tau)}(1) \ , \nonumber
\\
E^{(ab; \tau)} &\leftrightarrow& \sigma^{(ab; \tau)}(2) \ . 
\label{eq34}
\ea
The volume element for integration over all these variables is denoted
by ${\rm d}[\sigma]$. As a result, the ensemble average of the
generating function $Z$ takes the form
\ba
&&\overline{Z} = \int {\rm d} [\sigma] \exp \biggl [ - \sum_a {\rm
  trg} \biggl ( \frac{l^2}{8} \ (\sigma^{(aa)})^2 + \frac{l^2}{4} \
(\sigma^{(a((a+1))})^2 \nonumber \\
&&\qquad + \sum_{\tau} \biggl \{ \frac{l^2}{4} \ (\sigma^{(aa;
  \tau)})^2 + \frac{l^2}{4} \ (\sigma^{(a (a+1); \tau)})^2 \nonumber
\\
&&\qquad + \sum_{i=1}^2 \bigl ( \frac{N(-1,0)}{4} \ (\sigma^{((a-1) a;
  \tau)}(i))^2 + \frac{N(-1,1)}{4} \ (\sigma^{((a-1) (a+1);
  \tau)}(i))^2 \nonumber \\
&&\qquad + \frac{N(0,1)}{4} \ (\sigma^{(a (a+1); \tau)}(i))^2) \biggr
  \} \biggr )  - \ {\rm trg} \ {\rm tr} \ \ln \ {\bf N}(J) \biggr ] \
  .
\label{eq35}
\ea
Here ${\bf N}(J)$ is both, a graded matrix and a matrix in Hilbert
space with basis vectors $|a a \mu \rangle$ and $|a (a+1) \nu \rangle$.
In order to clearly display the structure of {\bf N}(J), we list
separately several types of matrix elements. There are two types of
diagonal matrix elements, the first type given by
\ba
&&\langle a a \mu | {\bf N}(J) | \ a a \nu \rangle = \biggl (E -
\lambda \sigma^{(aa)} \biggr ) \delta_{\mu \nu} \nonumber \\
&&\qquad - \sum_{\tau} \lambda \sigma^{(aa; \tau)} C^{(1
  \tau)}_{\mu \nu}(0) \ - w_{\mu \nu} \ .
\label{eq36}
\ea
As usual, we denote by $E$ half the sum of the energy arguments of both
Green's functions. The difference vanishes as both energies are equal.
The matrix $J$ denotes the source matrix. The second type of diagonal
matrix element is
\ba
&&\langle a (a+1) \mu | {\bf N}(J) | \ a (a+1) \nu \rangle = \biggl
(E \nonumber - \lambda \sigma^{(a (a+1))} \biggr ) \ \delta_{\mu \nu}
\\
&&\qquad - \sum_{\tau} \lambda \sigma^{(a (a+1); \tau)} C^{(1
  \tau)}_{\mu \nu}(1) \ .
\label{eq37}
\ea
The non--diagonal matrix elements of ${\bf N}(J)$ connect either
neighboring two--particle states,
\ba
&&\langle (a-1) a \mu | {\bf N}(J) | a a \nu \rangle = - v \ \delta_{\mu
\nu} \nonumber \\
&&\qquad - (1/2) \sum_{\tau} \lambda C^{(1 \tau)}_{\mu \nu}(-1,0;l)
(- i \sigma^{((a-1) a; \tau)}(1) + \sigma^{((a-1) a; \tau)}(2)) \ ,
\label{eq38}
\ea
\ba
&&\langle a a \mu | {\bf N}(J) | a (a+1) \nu \rangle =  - v \ \delta_{\mu
\nu} \nonumber \\
&&\qquad - (1/2) \sum_{\tau} \lambda C^{(1 \tau)}_{\mu \nu}(0,1;l)
(- i \sigma^{(a (a+1); \tau)}(1) + \sigma^{(a (a+1); \tau)}(2)) \ ,
\label{eq39}
\ea
\ba
&&\langle a a \mu | {\bf N}(J) | (a-1) a \nu \rangle = - v \ \delta_{\mu
  \nu} \nonumber \\
&&\qquad - (1/2) \sum_{\tau} \lambda C^{(1 \tau)}_{\mu \nu}(-1,0;r)
(i \sigma^{((a-1) a; \tau)}(1) + \sigma^{((a-1) a; \tau)}(2)) \ ,
\label{eq40}
\ea
\ba
&&\langle a (a+1) \mu | {\bf N}(J) | a a \nu \rangle = - v \ \delta_{\mu
\nu} \nonumber \\
&&\qquad - (1/2) \sum_{\tau} \lambda C^{(1 \tau)}_{\mu \nu}(0,1;r)
(i \sigma^{(a (a+1); \tau)}(1) + \sigma^{(a (a+1); \tau)}(2)) \ ,
\label{eq41}
\ea
or next--nearest states,
\ba
&&\langle (a-1) a \mu | {\bf N}(J) | a (a+1) \nu \rangle \nonumber \\
&&\qquad = - (1/2) \sum_{\tau} \lambda C^{(1 \tau)}_{\mu \nu}(-1,1;l)
\ \biggl (- i \sigma^{((a-1) (a+1); \tau)}(1) \nonumber \\
&&\qquad \qquad + \sigma^{((a-1) (a+1); \tau)}(2) \biggr ) \ ,
\label{eq42}
\ea
\ba
&&\langle (a+1) a \mu | {\bf N}(J) | a (a-1) \nu \rangle \nonumber \\
&&\qquad = - (1/2) \sum_{\tau} \lambda C^{(1 \tau)}_{\mu \nu}(-1,1;r)
\ \biggl (i \sigma^{((a-1) (a+1); \tau)}(1) \nonumber \\
&&\qquad \qquad + \sigma^{((a-1) (a+1); \tau)}(2) \biggr ) \ .
\label{eq43}
\ea
The matrix ${\bf N}(J)$ also contains the source terms $J$. These
depend upon the observable we are interested in. For the case of
localization properties of the two--point function, we define the
source terms in Section~\ref{non}. All other matrix elements of ${\bf
  N}(J)$ vanish. 

\section{Saddle--Point Equations}
\label{sad}

For an approximate evaluation of the averaged generating function
$\overline{Z}$, we follow standard procedure and use the saddle--point
approximation. This is suggested by the occurrence of large factors
$l^2$ and $N(i)$ in front of the quadratic terms in the exponent of
Eq.~(\ref{eq35}). We neglect $J$ which is infinitesimal. Following the
work of Ref.~\cite{iid}, we also neglect the hopping terms $v$. For
the single--electron problem with disorder, it was shown in
Ref.~\cite{iid} that by treating $v$ as a small perturbation of the
saddle--point solution, the standard non--linear sigma model for
localization is obtained. Technically speaking we assume that $v \ll
\lambda$. The matrix obtained from ${\bf N}(J)$ after all these
omissions is denoted by ${\bf N}'(0)$.

In the exponent of Eq.~(\ref{eq35}), we replace every $\sigma$--matrix
by $\sigma + \delta \sigma$. We collect the terms linear in $\delta
\sigma$ and put the coefficients multiplying the independent $\delta
\sigma$'s equal to zero. As a result, we find
\be
\sigma^{(aa)} = \frac{4}{l^2} \ {\rm tr} \biggl [ \biggl (
\frac{1}{{\bf N}'(0)} \biggr )_{(aa|aa)} \lambda \biggr ] \ .
\label{eq44}
\ee
The index $(aa|aa)$ indicates that the diagonal $aa$ box of the
inverse of ${\bf N}'(0)$ has to be taken. The trace is taken with
respect to the state indices $\mu, \nu$ in this box. Quite generally,
we use the expression ``box'' to denote the totality of Hilbert states
obtained by putting the two electrons in slices $a$ and $b$,
respectively, with $b = a$ or $b = (a+1)$. Using analogous notation,
we find
\ba
\sigma^{(aa; \tau)} &=& \frac{2}{l^2} \ {\rm tr} \biggl [ \biggl (
\frac{1}{{\bf N}'(0)} \biggr )_{(aa|aa)} \lambda C^{(1 \tau)}(0)
\biggr ] \ , \nonumber \\
\sigma^{(a(a+1))} &=& \frac{2}{l^2} \ {\rm tr} \biggl [ \biggl (
\frac{1}{{\bf N}'(0)} \biggr )_{(a(a+1)|a(a+1))} \lambda \biggr ] \ ,
\nonumber \\
\sigma^{(a(a+1); \tau)} &=& \frac{2}{l^2} \ {\rm tr} \biggl [ \biggl (
\frac{1}{{\bf N}'(0)} \biggr )_{(a(a+1)|a(a+1))} \lambda C^{(1
  \tau)}(1) \biggr ] \ .
\label{eq45}
\ea
For the terms which are not diagonal in the box indices, it is
convenient to introduce the matrices
\ba
\Sigma^{(ab; \tau)} &=& - \sigma^{(ab; \tau)}(1) - i \sigma^{(ab;
  \tau)}(2) \ , \nonumber \\
(\Sigma^{(ab; \tau)})^{\dagger} &=& - \sigma^{(ab; \tau)}(1) + i
\sigma^{(ab; \tau)}(2) \ .
\label{eq46}
\ea
The saddle--point equations for the index combination $((a-1)a)$ are
given by
\ba
\Sigma^{((a-1)a; \tau)} &=& \frac{2 i}{N(-1,0)} \ {\rm tr} \biggl [
\biggl ( \frac{1}{{\bf N}'(0)} \biggr )_{((a-1)a|aa)} \lambda C^{(1
  \tau)}(-1,0;r) \biggr ] \ , \nonumber \\
(\Sigma^{((a-1)a; \tau)})^{\dagger} &=& \frac{2 i}{N(-1,0)} \ {\rm tr}
\biggl [ \biggl ( \frac{1}{{\bf N}'(0)} \biggr )_{(aa|(a-1)a)} \lambda
C^{(1 \tau)}(-1,0;l) \biggr ] \ .
\label{eq47}
\ea
Corresponding equations are obtained for $((a-1)(a+1))$ and $a(a+1)$.
We test these results by adding Eq.~(\ref{eq44}) and the first of
Eqs.~(\ref{eq45}), and the second and third of Eqs.~(\ref{eq45}). With
the help of Eq.~(\ref{eq16}), this yields
\ba
\lambda \sigma^{(aa)} \delta_{\mu \nu} + \sum_{\tau} \lambda
\sigma^{(aa; \tau)} C^{(1 \tau)}_{\mu \nu}(0) &=& \frac{\lambda^2}{l}
\sum_{\rho \sigma} {\cal A}^{(0)}_{\mu \nu; \rho \sigma} \biggl (
\frac{1}{{\bf N}'(0)} \biggr )_{(aa|aa); \sigma \rho} \ , \nonumber \\
\lambda \sigma^{(a(a+1))} \delta_{\mu \nu} + \sum_{\tau} \lambda
\sigma^{(aa; \tau)} C^{(1 \tau)}_{\mu \nu}(1) &=& \frac{\lambda^2}{l}
\sum_{\rho \sigma} {\cal A}^{(1)}_{\mu \nu; \rho \sigma} \biggl (
\frac{1}{{\bf N}'(0)} \biggr )_{(aa|aa); \sigma \rho} \ . \nonumber \\
\label{eq48}
\ea
Similarly, we find
\ba
- \frac{i}{2} \sum_{\tau} \lambda C^{(1 \tau)}_{\mu \nu}(-1,0;l)
\Sigma^{((a-1)a; \tau)} &=& \frac{\lambda^2}{l} \sum_{\rho \sigma} 
{\cal A}^{(-1,0)}_{\mu \nu; \rho \sigma} \biggl ( \frac{1}{{\bf N}'(0)}
\biggr )_{((a-1)a|aa); \sigma \rho} \ , \nonumber \\
\frac{i}{2} \sum_{\tau} \lambda C^{(1 \tau)}_{\mu \nu}(-1,0;r)
(\Sigma^{((a-1)a; \tau)})^{\dagger} &=& \frac{\lambda^2}{l}
\sum_{\rho \sigma} {\cal A}^{(0,-1)}_{\mu \nu; \rho \sigma}
\nonumber \\
&& \times \biggl ( \frac{1}{{\bf N}'(0)} \biggr )_{((aa|(a-1)a);
  \sigma \rho} \ ,
\label{eq49}
\ea
and completely analogous relations for the index combinations
$((a-1)(a+1))$ and $(a(a+1))$. Eqs.~(\ref{eq48}) and (\ref{eq49})
provide a test for our calculation in the following sense. On the
left--hand sides of these equations, there occur the same linear
combinations of the $\sigma$--matrices as in the matrix ${\bf N}'(0)$,
see Eqs.~(\ref{eq36}) through (\ref{eq43}). On the right--hand side,
we encounter traces over the inverse of ${\bf N}'(0)$ times the
corresponding matrix element of the second moment of $H_0$. This is
exactly the same structure as encountered by Benet {\it et
  al.}~\cite{ben}. We note that using the orthonormality relations
of the coefficients $C^{(s \tau)}$, we can retrieve the original
saddle--point equations [ Eqs.~(\ref{eq44}) through (\ref{eq47}) ]
from Eqs.~(\ref{eq48}) and (\ref{eq49}). Thus, the latter set of
equations encapsulates the saddle--point conditions.

The structure of the saddle--point equations is displayed most
clearly when we introduce the following notation. For the matrix
${\bf N}'(0)$, we write
\be
\langle a b \mu | {\bf N}'(0) | a' b' \nu \rangle = (E \delta_{\mu
\nu} - w_{\mu \nu} \delta_{a b}) \delta_{a a'} \delta_{b b'} -
\langle a b \mu | X | a' b' \nu \rangle \ .
\label{eq50}
\ee
By definition, the matrix $X$ contains all the $\sigma$--matrices
which occur in ${\bf N}'(0)$. Needless to say that the possible index
combinations $(a,b)$ and $(a',b')$ are severely restricted by our
choice of admissible Hilbert vectors. Using this notation and the
definitions of the matrices ${\cal A}$ in Section~\ref{mom}, we can
cast the totality of saddle--point equations in the form
\be
\langle a b \mu | X | a' b' \nu \rangle = \sum_{\rho \sigma}
\overline{ \langle a b \mu | H_0 | a b \sigma \rangle \biggl (
\frac{1}{E - w -X} \biggr )_{(a b|a' b'); \sigma \rho} \langle a'
b' \rho | H_0 | a' b' \nu \rangle } \ .
\label{eq51}
\ee
Eq.~(\ref{eq51}) is the generalization of the Pastur equation to the
present problem.

\section{Solution of the Saddle--Point Equations}
\label{sol}

We follow Ref.~\cite{ben} and solve the saddle--point equations by
iteration. In Eq.~(\ref{eq51}), iteration generates a
continued--fraction expansion which we break off after $n$ steps where
$n$ is integer but arbitrary. The $n^{\rm th}$ element of the
continued--fraction expansion has the form
\be
\sum_{\rho \sigma} \overline{ \langle a b \mu | H_0 | a b \sigma
\rangle \biggl ( \frac{1}{E - w} \biggr )_{(a b|a' b'); \sigma \rho}
\langle a' b' \rho | H_0 | a' b' \nu \rangle } \ .
\label{eq52}
\ee
But $(E - w)$ is diagonal in the box indices $(ab)$ and $(a'b')$.
Therefore, expression~(\ref{eq52}) equals
\be
\sum_{\rho \sigma} \overline{ \langle a b \mu | H_0 | a b \sigma
\rangle \biggl ( \frac{1}{E - w} \biggr )_{(a b|a b); \sigma \rho}
\langle a b \rho | H_0 | a b \nu \rangle } \ \delta_{a a'} \delta_{b
b'} \ .
\label{eq53}
\ee
Using this result in the $(n-1)^{\rm st}$ element of the
continued--fraction expansion, we obtain a denominator which once
again is diagonal in the box indices $(ab)$ and $(a'b')$. Continuing
the argument all the way up to the first term, we conclude that all
contributions to $X$ which are not diagonal in the box indices,
vanish. Thus, we have for $i = 1,2$
\be
\sigma^{((a-1)a; \tau)}_{\rm sp}(i) = 0 \ ; \
\sigma^{((a-1)(a+1); \tau)}_{\rm sp}(i) = 0 \ ; \
\sigma^{(a(a+1); \tau)}_{\rm sp}(i) = 0 \ .
\label{eq54}
\ee
The index sp stands for saddle point. It follows that the
saddle--point equations separate into a set of $(2 K - 1)$ equations
for the box--diagonal elements of $X$. For $(ab) = (a(a+1))$ we have
$w = 0$, and the $n^{\rm th}$ element of the continued--fraction
expansion is given by
\ba
&&\sum_{\rho \sigma} \overline{ \langle a (a+1) \mu | H_0 | a (a+1)
\sigma \rangle E^{-1} \delta_{\sigma \rho} \langle a (a+1) \rho | H_0
| a (a+1) \nu \rangle } \nonumber \\
&&\qquad = \frac{\lambda^2}{l} \Lambda^{(0)}(1) \delta_{\mu \nu} \ .
\label{eq55}
\ea
The right--hand side of this equation follows from Eq.~(\ref{eq12})
and from the fact that $C^{(1 \tau)}_{\mu \nu}(-1)$ is traceless.
The fact that $E$ is multiplied by the unit matrix $\delta_{\mu \nu}$
implies the same statement for the $n^{\rm th}$ element of the
continued--fraction expansion. Again, we can continue the argument
all the way up to the first term and conclude that
\be
\sigma^{(a (a+1); \tau)}_{\rm sp} = 0 \ .
\label{eq56}
\ee
Therefore, the only remaining matrix is $\sigma^{(a(a+1))}$. We use
Eqs.~(\ref{eq7a}) and (\ref{eq10}), the fact that in the latter only
the term with $s = 0$ is relevant, and the fact that $C^{(0)}_{\mu \nu}
= \delta_{\mu \nu}$. It follows that $\sigma^{(a(a+1))}$ obeys the
standard saddle--point equation
\be
\sigma^{(a(a+1))}_{\rm sp} = \frac{\lambda}{E - \lambda
\sigma^{(a(a+1))}_{\rm sp}} \ .
\label{eq57}
\ee
As usual, the solution is obtained in two steps. First, we determine
the diagonal elements $\sigma^{(a(a+1))}_{\rm d}$ which obey
Eq.~(\ref{eq57}). For $|E| \leq 2 \lambda$, these are given by
\be
\sigma^{(a(a+1))}_{\rm d} = \frac{E}{2 \lambda} \pm i \Delta_{1}(E) \,
\label{eq58}
\ee
where we have defined
\be
\Delta_{1}(E) = \sqrt{1 - \biggl ( \frac{E}{2 \lambda} \biggr )^2 } \ .
\label{eq58a}
\ee
We observe that the quantity $\Delta_{1}(E)$ is proportional to the
average level density $\rho_{{\rm sp} 1}(E)$. Eq.~(\ref{eq58a}) shows
that $\rho_{{\rm sp} 1}(E)$ has the shape of a semicircle with radius
$2 \lambda$. The invariance of the saddle--point equation
Eq.~(\ref{eq58}) under pseudounitary graded transformations $T$
implies that the saddle--point manifold is given by~\cite{ver}
\ba
\sigma^{(a(a+1))}_{\rm sp} = T^{(a (a+1))} \ [ \ \frac{E}{2 \lambda} -
i \ \Delta_{1}(E) \ L \ ] \ ( T^{(a (a+1))} )^{-1} \nonumber \\
= \frac{E}{2 \lambda} - i \ \Delta_{1}(E) \ T^{(a (a+1))} L ( T^{(a
  (a+1))} )^{-1} \ .
\label{eq59}
\ea

For the index combination $(aa)$, the saddle--point equation contains
the two--body interaction $w$. We are interested in comparing the cases
$w = 0$ and $w \neq 0$ and, therefore, treat both cases. For $w = 0$,
we conclude as before that 
\be
\sigma^{(a a; \tau)}_{\rm sp} = 0 \ {\rm for} \ w = 0 \ .
\label{eq60}
\ee
For $\sigma^{(aa)}$, we obtain the same saddle--point equation
[Eq.~(\ref{eq57})] as for $\sigma^{(a(a+1))}$. Eq.~(\ref{eq44}) and
the second of Eqs.~(\ref{eq45}) actually differ by a factor 2. The
factor disappears, however, when we take the traces over $\delta_{\rho
\sigma}$ because the dimensions $N_{aa}$ and $N_{a(a+1)}$ differ by
that same factor. The solution is accordingly given by
\ba
\sigma^{(a a)}_{\rm sp} &=& \ T^{(a a)} \ [ \ \frac{E}{2 \lambda} - i
\ \sqrt{1 - \biggl ( \frac{E}{2 \lambda} \biggr )^2 } \ L \ ] \ (
T^{(a a)} )^{-1} \nonumber \\
&=& \ \frac{E}{2 \lambda} - i \ \sqrt{1 - \biggl ( \frac{E}{2 \lambda}
  \biggr )^2 } \ T^{(a a)} \ L \ ( T^{(a a)} )^{-1} \ {\rm for} \ w =
0 \ .
\label{eq61}
\ea
For $w \neq 0$, we encounter a novel situation because $w_{\mu \nu}$
is a genuine matrix and the arguments used above do not apply. The
$n^{\rm th}$ element of the continued--fraction expansion has the form
\be
\sum_{\rho \sigma} \overline{ \langle a a \mu | H_0 | a a \sigma
\rangle \biggl ( \frac{1}{E - w} \biggr )_{(a a|a a); \sigma \rho}
\langle a a \rho | H_0 | a a \nu \rangle } \ .
\label{eq62}
\ee
The second moment of $H_0$ in expression~(\ref{eq62}) is proportional
to the matrix ${\cal A}^{(0)}$ in Eq.~(\ref{eq7b}) for which we use
the eigenvalue expansion Eq.~(\ref{eq16}). We first consider the terms
with $s = 1$ for which the sum over $(\rho, \sigma)$ takes the form
$\sum_{\rho \sigma} C^{(1 \tau)}_{\rho \sigma}(0) (1/[ E - w ]
)_{(a a|a a); \sigma \rho}$ with $C^{(1 \tau)}_{\mu \nu}(0) \propto
\langle a a \mu | \alpha^{\dagger}_m \alpha_n | a a \nu \rangle$
(where $m \neq n$) or $C^{(1 \tau)}_{\mu \nu}(0) \propto \langle a a
\mu | (\alpha^{\dagger}_m \alpha_m - (1/l) \sum_n \alpha^{\dagger}_n
\alpha_n ) |  a a \nu \rangle$. We write $M_{\mu \nu} = ((E -
w)^{-1})_{\mu \nu}$. For $C^{(1 \tau)}_{\mu \nu}(0) \propto \langle a
a \mu | \alpha^{\dagger}_m \alpha_n | a a \nu \rangle$, we find
$\sum_{\rho \sigma} C^{(1 \tau)}_{\rho \sigma}(0) M_{\mu \nu} = (4/l^2)
\sum_j M_{n j m j}$ where we have returned to the original notation in
Eq.~(\ref{eq5}). Expanding $M$ in powers of $w$ and taking the linear
term, we note that the matrix element $w_{n j m j}$ vanishes for $m
\neq n$. This can easily be seen by using in each slice periodic
boundary conditions in longitudinal direction and follows from momentum
conservation. The same statement applies to all higher moments of $w$.
We conclude that $\sum_{\rho \sigma} \langle a a \rho |
\alpha^{\dagger}_m \alpha_n | a a \sigma \rangle M_{\sigma \rho} = 0$.
We turn to $C^{(1 \tau)}_{\mu \nu}(0) \propto \langle a a \mu |
(\alpha^{\dagger}_m \alpha_m - (1/l) \sum_n \alpha^{\dagger}_n
\alpha_n ) |  a a \nu \rangle$ and find that $\sum_{\rho \sigma} C^{(1
  \tau)}_{\rho \sigma}(0) M_{\mu \nu}$ is proportional to $(4/l^2) (
\sum_j M_{m j m j} - (1/l) \sum_{j n} M_{n j n j} )$. We again
consider first the term linear in $w$. Although we do not expect the
condition $w_{m j m j} = w_{n j n j}$ to hold for arbitrary values of
$m$ and $n$, we believe that averaging over $j$ leads to an
approximate fulfillment of the condition $\sum_j w_{m j m j} =
\sum_{j} w_{n j n j}$ for arbitrary values of $m$ and $n$. This is
expected to be true {\it a fortiori} for higher powers of $w$. Thus,
in very good approximation the terms with $s = 1$ do not contribute to
the expression~(\ref{eq62}), the argument applies equally to all terms
in the continued--fraction expansion, and we obtain
\be
\sigma^{(aa; \tau)}_{\rm sp} \approx 0 \ {\rm for} \ w \neq 0 \ .
\label{eq63}
\ee

We are left with $\sigma^{(aa)}$ which obeys the saddle--point equation
\be
\sigma^{(aa)}_{\rm sp} = N_{aa}^{-1} \ {\rm tr} [ ( \frac{\lambda}{E
- w - \lambda \sigma^{(aa)}_{\rm sp}} )_{\mu \nu} ] \ .
\label{eq64}
\ee
This is easily verified by comparing the continued--fraction expansion
of Eq.~(\ref{eq64}) with that of Eq.~(\ref{eq53}) for $a = b$ after we
drop in the latter the terms with $s = 1$. We recall that $N_{aa} =
l(l-1)/2$. The matrix $w$ is Hermitean. We denote the eigenvalues of
$w$ by $\varepsilon_j$ with $j = 1,\ldots,N_{aa}$. Without loss of
generality, we choose $\varepsilon_1 \leq \varepsilon_2 \leq \ldots
\leq \varepsilon_{N_{aa}}$. The saddle--point equation Eq.~(\ref{eq64})
takes the form
\be
\sigma^{(aa)}_{\rm sp} = N_{aa}^{-1} \sum_j \frac{\lambda}{E -
\varepsilon_j - \lambda \sigma^{(aa)}_{\rm sp}} \ .
\label{eq65}
\ee
As is the case for $w = 0$, this equation is invariant under all
graded pseudounitary transformations $T^{(aa)}$. Therefore, we first
determine the scalar solutions $\sigma^{(aa)}_{\rm d}$ and from there
construct the saddle--point manifold as in Eq.~(\ref{eq61}).
Eq.~(\ref{eq65}) can be written as a polynomial of order $N_{aa}$ in
$\sigma^{(aa)}_{\rm d}$ and, therefore, possesses $N_{aa}$ real or
complex solutions. We show that of these, $N_{aa} - 2$ are always real.
The remaining two solutions are either real or complex conjugate to
each other. The positive imaginary part determines the level density
$\rho_{{\rm sp} 0}(E)$ in the box $(aa)$. Therefore, the spectrum
extends over that energy interval where two complex conjugate
solutions of Eq.~(\ref{eq65}) exist. We determine the spectrum
geometrically.

As a preparatory step, we return to Eq.~(\ref{eq57}). With $z =
E/\lambda$ and $\tau = z - \sigma^{a(a+1)}_{\rm d}$, we write this
equation in the form
\be
z - \tau = 1/\tau \ .
\label{eq66}
\ee
The right--hand side is a hyperbola with asymptotes along the abscissa
and the ordinate. The left--hand side represents a bundle of straight
lines which intersect the abscissa at an angle of $- \pi / 4$. For
sufficiently large values of $|E|$, each of these straight lines
intersects the hyperbola twice while for small values of $|E|$, there
are no points of intersection. The end points of the spectrum coincide
with the points where a straight line osculates the hyperbola, i.e.,
where the derivative of the hyperbola equals $-1$. This is the case at
$\tau = \pm 1$ or, with Eq.~(\ref{eq66}), where $z = \pm 2$ and, thus,
$E = \pm 2 \lambda$. This agrees with Eq.~(\ref{eq58}).

We apply the same consideration to Eq.(\ref{eq65}) and again use $z =
E/\lambda$ and $\tau = z - \sigma^{(aa)}_{\rm d}$. Then,
\be
z - \tau = N_{aa}^{-1} \sum_j \frac{1}{\tau - \varepsilon_j/\lambda} \
.
\label{eq67}
\ee
Each term on the right--hand side of this equation is a hyperbola with
asymptotes along the abscissa and along a straight line parallel to
the ordinate which intersects the abscissa at $\tau = \varepsilon_j /
\lambda$. The left--hand side again represents a bundle of straight
lines which intersect the abscissa at an angle of $- \pi / 4$.
Figure~1 represents the situation schematically. We see that for
sufficiently large values of $|E|$, each straight line intersects the
sum of the hyperbolas exactly $N_{aa}$ times, giving rise to $N_{aa}$
real solutions. For sufficiently small values of $|E|$, the number of
intersections is $N_{aa} - 2$. The spectrum extends continuously from
a point $E_1$ to the left of $\varepsilon_1$ to a point $E_2$ to the
right of $\varepsilon_{N_{aa}}$. Again, the energies $E_{1,2}$ are
defined in terms of $\tau_1$ and $\tau_2$. The latter are those points
where one of the straight lines $z - \tau$ osculates the sum of the
hyperbolas, i.e., where the gradient of the latter is $-1$. With
increasing strength of the two--body interaction $w$, the difference
$\varepsilon_{N{aa}} - \varepsilon_1$ grows monotonically and so does,
therefore, the difference $E_2 - E_1$. This difference gives the width
of the spectrum in the box $(aa)$. Thus, the Wigner semicircle
appearing in Eq.~(\ref{eq58}) is gradually deformed by becoming wider
and flatter (the area remains the same).

We supplement these general considerations by calculating position and
width of the spectrum perturbatively for small values of $w$. It is
easily seen that the first--order term only shifts the center of the
semicircle from zero to ${\rm tr} (w) / N_{aa}$, without changing
either radius or shape of the semicircle. We accordingly define $E_0 =
{\rm tr} (w) / N_{aa} $ and $z_0 = E_0 / \lambda$. Introducing the
normalized variance of $w$ as $U^2 = \{ (1/N_{aa}) {\rm tr} (w^2) - [
(1/N_{aa}) {\rm tr} (w) ]^2 \} / \lambda^2$ and expanding the
right--hand side of Eq.~(\ref{eq67}) in powers of $w$, we get
\be
z - z_0 - \tau = \frac{1}{\tau} + \frac{U^2}{\tau^3} + \ldots \ .
\label{eq68}
\ee
To lowest order in $U^2$, the points $\tau_1$ and $\tau_2$ are given
by $\tau_{1,2} = \pm (1 + (3/2) U^2 )$. The end points $E_{1,2}$ of
the spectrum are accordingly given by
\be
E_{1,2} = E_0 \pm (2 + U^2) \lambda \ . 
\label{eq69}
\ee
This result clearly shows the widening of the spectrum.

How strong a two--body interaction is needed to yield a sizeable
effect on the localization length? A qualitative change of the
spectrum in the box $(aa)$ and, thus, in localization occurs whenever
$E_{1,2}$ and, thus, $E_0$ and $U^2 / \lambda$ are of the order of
$\lambda$, the width of the unperturbed two--body spectrum in box
$(aa)$. In order to be independent of the sign of $w$, we focus
attention on one contribution to $U^2$, i.e., the term $(1/N_{aa})
{\rm tr} (w^2) / \lambda^2$ which must then be of order unity. For
realistic two--body interactions, the expression ${\rm tr}(w^2)$ may
not exist. Therefore, we use completeness and write $(1/N_{aa}) {\rm
  tr} (w^2)$ in the form $\langle a a \mu | w^2 | a a \mu \rangle_{\rm
  av}$ where the index av stands for an average over the states
labelled $\mu$. We obtain $\langle a a \mu | w^2 | a a \mu
\rangle_{\rm av} = \lambda^2$ . To eliminate $\lambda$, we use
Eq.~(\ref{eq2}) which yields $\lambda^2 = \sum_j
\overline{(h^{(a)}_{ij})^2}$. We identify the matrix elements
$h^{(a)}_{ij}$ with the matrix elements $\langle a i | V_{\rm imp} | a
j \rangle$ of the impurity potential $V_{\rm imp}$. We do so because
upon discretization the kinetic (potential) terms of a continuum model
turn into hopping matrix elements (local energies, respectively). We
replace the ensemble average by a running average over the
single--particle states, use completeness, and obtain
\be
\langle a a \mu | w^2 | a a \mu \rangle_{\rm av} \geq \langle a i |
(V_{\rm imp})^2 | a i \rangle_{\rm av} \ .
\label{crit}
\ee
Whenever the criterion~(\ref{crit}) is met, the two--body interaction
will change the localization length qualitatively. Quantitative
changes set in for smaller values of $w^2$, of course. Because of
the normalization of the wave functions, the criterion~(\ref{crit}) is
essentially independent of the size of the slices. This is physically
reasonable. Moreover, the criterion~(\ref{crit}) can easily be checked
in concrete cases. Since the screened Coulomb interaction is more or
less fixed, the criterion~(\ref{crit}) effectively establishes an
upper bound on the strength of the impurity potential for interaction
effects to be relevant. The criterion~(\ref{crit}) has only
qualitative significance because it is based upon the saddle--point
approximation. When we identify the left--hand side with the
mean--square matrix element $U^2$ in Eq.~(\ref{0}) and the left--hand
side with the bandwidth $B^2$ we see the close correspndence between
our criterion and the result Eq.~(\ref{0}). We point out, however,
that on the quantitative level, our result differs from Eq.~(\ref{0}).
This is discussed in Sections~\ref{loop1} and \ref{disc}.

Returning to the general case, we write $\sigma^{(aa)}_{\rm d}$ in the
form
\be
\sigma^{(aa)}_{\rm d} = a(E) \pm i \Delta_{0}(E)
\label{eq70}
\ee
where $\Delta_{0}(E) > 0$ for $E_1 < E < E_2$ is proportional to the
spectral density $\rho_{{\rm sp} 0}(E)$. The general form of the
saddle--point solution is then
\ba
\sigma^{(aa)}_{\rm sp} = T^{(aa)} \ [ \ a(E) - i \Delta_{0}(E) \ L
\ ] \ (T^{(aa)})^{-1} \nonumber \\
= a(E) - i \Delta_{0}(E) \ T^{(aa)} L (T^{(aa)})^{-1} \ .
\label{eq71}
\ea
The essential difference between the saddle--point solutions in
Eqs.~(\ref{eq71}) and (\ref{eq59}) lies in the difference between
$\Delta_{0}(E)$ and $\Delta_{1}(E)$, i.e, in the different spectral
densities $\rho_{{\rm sp} 0}(E)$ and $\rho_{{\rm sp} 1}(E)$. Only for
$w = 0$ do we have $\Delta_{0}(E) = \Delta_{1}(E)$.

\section{Non--Linear Sigma Model. Localization \\ Properties}
\label{non}

We now expand the effective Lagrangean in the exponent of
Eq.~(\ref{eq35}) around the saddle--point solutions obtained in
Section~\ref{sol}. To this end, we write for $i = 1,2$
\ba
\sigma^{(aa)} &=& \sigma^{(aa)}_{\rm sp} + T^{(aa)} \delta P^{(aa)}
(T^{(aa)})^{-1} \ , \nonumber \\
\sigma^{(a(a+1))} &=& \sigma^{(a(a+1))}_{\rm sp} + T^{(a(a+1)} \delta
P^{(a(a+1)} (T^{(a(a+1))})^{-1} \ , \nonumber \\
\sigma^{(aa; \tau)} &=& T^{(aa)} \delta \sigma^{(aa; \tau)}
(T^{(aa)})^{-1} \ , \nonumber \\
\sigma^{(a(a+1); \tau)} &=& T^{(a(a+1)} \delta \sigma^{(a(a+1);
  \tau)} (T^{(a(a+1))})^{-1} \ , 
\nonumber \\
\sigma^{((a-1)a; \tau)}(i) &=& T^{((a-1)a)} \delta
\sigma^{((a-1)a; \tau)}(i) (T^{(aa)})^{-1} \ , \nonumber \\
\sigma^{((a-1)(a+1); \tau)}(i) &=& T^{((a-1)a)} \delta
\sigma^{((a-1)(a+1); \tau)}(i) (T^{(a(a+1))})^{-1} \ , \nonumber \\ 
\sigma^{(a(a+1); \tau)}(i) &=& T^{(aa)} \delta \sigma^{(a(a+1);
  \tau)}(i) (T^{(a(a+1))})^{-1} \ .
\label{eq72}
\ea
The quantities $\delta P^{(aa)}$ and $\delta P^{(a(a+1)}$ are
block--diagonal in a representation of the supermatrices where the
first (second) block of dimension two corresponds to the first (the
second) Green's function, respectively, see Ref.~\cite{ver}. In
writing Eqs.~(\ref{eq72}), we have used the freedom to multiply each
$\delta \sigma$ and each $\delta P$ from the left (right) by the
corresponding matrix $T$ ($T^{-1}$, respectively). The volume of
integration ${\rm d}[\sigma]$ in Eq.~(\ref{eq35}) changes into the
product of the integration measures for all the $T$'s, $\delta P$'s,
and $\delta \sigma$'s. Substituting Eqs.~(\ref{eq72}) into
Eq.~(\ref{eq35}) and expanding the exponent in powers of the $\delta
P$'s and $\delta \sigma$'s, we generate terms of order $0, 2$ and $t
\geq 3$. (The linear terms vanish approximately because of the
saddle--point condition, approximately because of Eq.~(\ref{eq63})).
In the present Section, we focus attention on the terms of order
zero. The remaining terms give rise to the loop expansion. This
expansion is investigated in Sections~\ref{loop} and \ref{loop1}
below.

In addition to the $\delta \sigma$'s and $\delta P$'s, ${\bf N}(J)$
also contains $v$ and $J$, and we must clarify how to handle
contributions arising from these terms. We first address the choice of
$J$ which is needed for investigating localization. We consider the
average two--point function $C(n)$ defined in Eq.~(\ref{a1}). If the
system is localized, this quantity decays for $n \gg 1$ exponentially
with increasing $n$. The exponential decay is governed by the
localization length. In the expression~(\ref{a1}), the label $a$ need
not be equal to $1$ nor do we put $(a+n+1)$ equal to $K$, the total
number of slices. In fact, choosing $1 \ll a \ll (a+n) \ll K$ is
expected to display the localization properties without edge effects.
Instead of the Green's functions in expression~(\ref{a1}), we might
have considered another expression where the bra state (ket state) is
chosen as $\langle a a|$ (as $| (a+n)(a+n)|\rangle$, respectively).
Localization properties should not depend upon this choice, and this
is indeed what we shall find. To generate the expression~(\ref{a1})
from the source terms, we choose the matrix $J$ to have only the
following non--zero matrix elements.
\ba
\langle a (a+1) \mu | J | (a+n) (a+n+1) \nu \rangle = \delta_{\mu
  \mu_0} \delta_{\nu \nu_0} k_{\alpha \beta} j_{\alpha \beta}
\nonumber \\ 
\langle (a+n) (a+n+1) \nu | J | a (a+1) \mu \rangle = \delta_{\mu
  \mu_0} \delta_{\nu \nu_0} k_{\alpha \beta} j_{\alpha \beta} \ .
\label{a4}
\ea
Here $k$ is a four--by--four graded diagonal matrix with diagonal
matrix elements $+1$ ($-1$) in the Boson--Boson block (the
Fermion--Fermion block, respectively), and $j$ is a four--by--four
graded diagonal matrix with diagonal matrix elements $j_1$ ($j_2$) for
the retarded (advanced) Green's function, respectively. The
term~(\ref{a1}) is obtained by differentiating $\overline{Z}$ once
with respect to both $j_1$ and $j_2$ at $j_1 = 0 = j_2$.

In order to illuminate the role of the hopping terms $v$, it is
instructive to display the structure of the source terms for the 
case of the saddle--point solution, i.e., when all the $\delta
\sigma$'s and $\delta P$'s are put equal to zero. We define
\ba
g^{(a (a+1))}_{\mu \nu} &=& T^{(a (a+1))} \ g^{(a (a+1))}_{{\rm d}; \mu
  \nu} \ ( T^{(a (a+1))})^{-1} \nonumber \\
&=& T^{(a (a+1))} \ \delta_{\mu \nu} \ [ \frac{E}{2 \lambda}
- i \Delta_1 L ]^{-1} \ ( T^{(a (a+1))})^{-1} \ , \nonumber \\
g^{(aa)}_{\mu \nu} &=& T^{(aa)} \ g^{(aa)}_{{\rm d}; \mu \nu} \ (
T^{(aa)})^{-1} \nonumber \\
&=& T^{(aa)} \ ( [ \frac{E}{2 \lambda} - a(E) - i \Delta_0 L ) - w
]^{-1} )_{\mu \nu} \ ( T^{(aa)})^{-1} \ .
\label{a5}
\ea
We put $v = 0$ and expand ${\rm tr} \ {\rm trg} \ \ln {\bf N}(J)$ in
powers of $J$, keeping terms of orders $1$ and $2$ which we denote by
$(0,1)$ and $(0,2)$, respectively. We expand the exponential in powers
of $J$ and keep the terms of second order, given by $(0,1)^2$ and by
$(0,2)$. The term $(0,1)$ actually vanishes and the term $(0,2)$ is
proportional to
\ba
&&{\rm trg} \sum_{\mu \nu \rho \sigma} \ g^{(a (a+1))}_{\mu \nu}
\ \langle a (a+1) \nu | J | (a+n) (a+n+1) \rho \rangle \nonumber \\
&& \qquad \qquad \times g^{((a+n) (a+n+1))}_{\rho \sigma} \ \langle
(a+n) (a+n+1) \sigma | J | a (a+1) \mu \rangle \ . 
\label{a6}
\ea
This is the standard term used in the calculation of localization
properties using the non--linear sigma model~\cite{zir,mir}. It is
easy to see that there are non--vanishing contributions from higher
orders in $v/\lambda$. Since $v \ll \lambda$, such contributions can
be omitted.

It remains to work out the terms proportional to $(v / \lambda)^2$.
We have already calculated the source terms and, therefore, put $J =
0$. The matrix ${\bf N}_{\rm sp}(0)$ has the following non--zero
matrix elements.
\ba
&&\langle a a \mu | {\bf N}_{\rm sp}(0) | \ a a \nu \rangle = \biggl
(E - \lambda \sigma^{(aa)}_{\rm sp} \biggr ) \delta_{\mu \nu} - w_{\mu
  \nu} \ , \nonumber \\
&&\langle a (a+1) \mu | {\bf N}_{\rm sp}(0) | \ a (a+1) \nu \rangle =
\biggl (E - \lambda \sigma^{(a (a+1))} \biggr ) \ \delta_{\mu \nu} \ ,
\nonumber \\
&&\langle (a-1) a \mu | {\bf N}_{\rm sp}(0) | a a \nu \rangle =
\langle a a \mu | {\bf N}_{\rm sp}(0) | (a-1) a \nu \rangle = - v \
\delta_{\mu \nu} \ , \nonumber \\
&&\langle a a \mu | {\bf N}_{\rm sp}(0) | a (a+1) \nu \rangle =
\langle a (a+1) \mu | {\bf N}_{\rm sp}(0) | a a \nu \rangle = - v \
\delta_{\mu \nu} \ .
\label{eq73}
\ea
We use the fact that $\sigma^{(aa)}_{\rm sp}$ and
$\sigma^{(a(a+1))}_{\rm sp}$ are solutions of the saddle--point
Eqs.~(\ref{eq65}) and (\ref{eq57}). Moreover, both the matrix $v$ and
$\sigma^{(a(a+1))}_{\rm sp}$ are proportional to $\delta_{\mu \nu}$.
This yields
\ba
&&- {\rm trg} \ {\rm tr} \ln {\bf N}_{\rm sp}(0) = + (v/\lambda)^2
\sum_a [ ( N_{aa} + N_{a(a+1)} ) / 2 ] \nonumber \\
&&\qquad \qquad \times [ {\rm trg} ( \sigma^{(aa)}_{\rm sp}
\sigma^{(a(a+1))}_{\rm sp} ) + {\rm trg} (\sigma^{(a(a+1)}_{\rm sp}
\sigma^{((a+1)(a+1))}_{\rm sp} ) ] \ . 
\label{eq74}
\ea
The right--hand side of Eq.~(\ref{eq74}) contains the source terms
(proportional to powers of $J$) and, most importantly, the term
responsible for electron transport through the wire. With
Eqs.~(\ref{eq59}) and (\ref{eq71}) and the explicit values of $N_{aa}$
and $N_{a(a+1)}$, this last term has the form
\ba
&& + (v/\lambda)^2 (3 l^2 / 4 ) \Delta_{0} \Delta_{1} \biggl ( \sum_a
\ {\rm trg} ( T^{(aa)} L (T^{(aa)})^{-1} T^{(a(a+1))} L
(T^{(a(a+1))})^{-1} ) \nonumber \\
&&+ \sum_a \ {\rm trg} ( T^{(a(a+1))} L (T^{(a(a+1))})^{-1}
T^{((a+1)(a+1))} L (T^{((a+1)(a+1))})^{-1} ) \biggr ) \ .
\label{eq75}
\ea
We recall that we are interested in comparing electron transport
through the wire for the cases $w = 0$ and $w \neq 0$. We note that
in the formulation of Eq.~(\ref{eq75}), the only difference between
the two cases lies in the difference of the values of $\Delta_{0}(E)$.
For $w = 0$ we have $\Delta_{0}(E) = \Delta_{1}(E) = \sqrt{1 - (E/2
  \lambda)^2}$ while this equality is violated for $w \neq 0$.

Eq.~(\ref{eq75}) establishes a non--linear sigma model for the
transport of two electrons through a quasi one--dimensional wire. This
model is equivalent to the model for the transport of a single
electron through the same wire studied in Ref.~\cite{iid}. To see
this, we recall that the model of Ref.~\cite{iid} also considered the
wire as divided into $K'$ slices numbered $j = 1,\ldots,K'$. We put
$K' = 2K - 1$ and map the $2K - 1$ boxes labelled $(aa)$ and $a(a+1)$
of the two--electron problem onto the $K'$ slices of the
single--electron problem of Ref.~\cite{iid} by putting $j = 2a -1$ for
the boxes $(aa)$ and $j = 2a$ for the boxes $(a(a+1)$. Then, the terms
in Eq.~(\ref{eq75}) take the form
\be
+ (v/\lambda)^2 (3 l^2 / 4 ) \Delta_{0} \Delta_{1} \sum_{j = 1}^{K'}
\ {\rm trg} ( T^{(j)} L (T^{(j)})^{-1} T^{(j+1)} L (T^{(j+1)})^{-1} )
\ .
\label{eq76}
\ee
This is exactly the form of the non--linear sigma model derived for
the transport of a single electron through a disordered wire in
Ref.~\cite{iid}. A similar correspondence can be established for the
source terms. The localization properties of the non--linear sigma
model Eq.~(\ref{eq76}) for single electron transport were extensively
studied in Refs.~\cite{zir,mir}. The correspondence between the models
in Eqs.~(\ref{eq76}) and (\ref{eq75}) makes it possible to use these
results for the discussion of the localization properties of the model
of two interacting electrons in Eq.~(\ref{eq75}). All we have to do is
to transcribe the results of Refs.~\cite{zir,mir} into the present
framework.

In Refs.~\cite{zir,mir}, the continuum limit was taken by letting the
length of each slice go to zero and the number $K'$ of slices go to
infinity. The same must be done in the framework of Eq.~(\ref{eq75}).
We take this step in Section~\ref{loop1} below. At the moment, it
suffices to observe that the localization length is directly
proportional to the coefficient $(v/\lambda)^2 (3 l^2 / 4 ) \Delta_{0}
\Delta_{1}$ appearing in Eq.~(\ref{eq76}). This statement then holds
likewise for Eq.~(\ref{eq75}). We recall that $\Delta_{0}$ and
$\Delta_{1}$ are proportional to the spectral densities $\rho_{{\rm
    sp} 0}(E)$ and $\rho_{{\rm sp} 1}(E)$, respectively, as determined
by the saddle--point condition. We accordingly define the
saddle--point approximation $T_{\rm sp}$ to the transport coefficient
(sometimes also referred to as the microscopic dimensionless
conductance) as
\be
T_{\rm sp} = 2 \pi \ \rho_{{\rm sp} 0}(E) \ v^2 \ \rho_{{\rm sp} 1}(E)
\ .
\label{eq76a}
\ee
This definition is patterned after the general definition of transport
coefficients for stochastic quantum problems in Ref.~\cite{agassi}.
Transport is mediated by the strength $v^2$ of the hopping matrix
elements connecting neighboring groups of states. The dimensionless
transport coefficient is symmetric with respect to the interchange of
the two boxes $(a a)$ and $(a (a+1))$. Transport is possible only if
the product $\rho_{{\rm sp} 0}(E) \ \rho_{{\rm sp} 1}(E)$ differs from
zero. Our result~(\ref{eq77}) is subject to this constraint. Let
$\Delta E_1 \ (\Delta E_0(w))$ denote the energy interval where
$\rho_{{\rm sp} 1}(E) \neq 0 \ (\rho_{{\rm sp} 0}(E) \neq 0$,
respectively). We obviously have $\Delta E_0(w = 0)) = \Delta_1$.
Typically we expect that for $w \neq 0$ the interval $\Delta E_0(w))$
is both shifted and widened in comparison with $\Delta E_1$, the
widening outweighing the shift. In this case, the energy interval
where electron transport is possible is given by $\Delta E_0(w = 0)) =
\Delta E_1$.

The ratio of the localization lengths $\zeta(w \neq 0)$ for
non--vanishing two--body interaction and $\zeta(w = 0)$ for vanishing
two--body interaction, is accordingly given by 
\be
\frac{\zeta(w \neq 0)}{\zeta(w = 0)} = \frac{\rho_{{\rm sp} 0}(E)}
{\rho_{{\rm sp} 1}(E)} \ .  
\label{eq77}
\ee
Eq.~(\ref{eq77}) is valid within the same energy interval as the
result~(\ref{eq76a}) for the transport coefficient. The
ratio~(\ref{eq77}) should typically be larger (smaller) than unity at 
the edges (in the middle) of the spectrum.

While physically very plausible, the form of Eq.~(\ref{eq75}) has one
obvious flaw. The transport coefficient is proportional to the
approximate level densities as determined by the saddle--point
approximation rather than to the exact level densities. And we know
for sure that at least $\rho_{{\rm sp} 1}(E)$ (which has the shape of
a semicircle) differs from the exact level density. The latter is the
level density of two non--interacting electrons each subject to a
random Hamiltonian and, therefore, given by the convolution of two
one--body densities of semicircle shape. This yields a non--zero
density in the interval $[ -4 \lambda, 4 \lambda]$ the shape of which
is intermittent between a semicircle and a Gaussian. We conclude that
Eq.~(\ref{eq75}) cannot possibly be the exact answer. This is why we
turn now to a study of the loop expansion (the expansion of the
integrand of the generating function around the saddle--point
solution).

\section{Loop Expansion: Terms of order zero in $v$}
\label{loop}

We investigate the terms in the loop expansion of $\overline{Z}$. We
recall that this expansion is generated by expanding the exponent in
Eq.~(\ref{eq35}) in powers of the $\delta \sigma$'s and $\delta P$'s,
by expanding the exponential containing terms of higher order than the
second in a Taylor series, and by carrying out the resulting Gaussian
integrations.

Expanding the quadratic terms in the exponent in Eq.~(\ref{eq35}) in
powers of the $\delta \sigma$'s and $\delta P$'s is trivial. Terms of
zeroth order vanish. Terms of first order cancel against those
stemming from the expansion of ${\rm tr} \ {\rm trg}\ \ln {\bf N}(J)$.
The terms of second order are
\ba
&&- \sum_a {\rm trg} \biggl ( \frac{l^2}{8} \ (\delta P^{(aa)})^2 +
\frac{l^2}{4} \ (\delta P^{(a((a+1))})^2 \nonumber \\
&&\qquad + \sum_{\tau} \biggl \{ \frac{l^2}{4} \ (\delta \sigma^{(aa;
  \tau)})^2 + \frac{l^2}{4} \ (\delta \sigma^{(a (a+1); \tau)})^2
\nonumber \\
&&\qquad + \sum_{i=1}^2 \bigl ( \frac{N(-1,0)}{4} \ (\delta
\sigma^{((a-1) a; \tau)}(i))^2 + \frac{N(-1,1)}{4} \ (\delta
\sigma^{((a-1) (a+1); \tau)}(i))^2 \nonumber \\
&&\qquad + \frac{N(0,1)}{4} \ (\delta \sigma^{(a (a+1); \tau)}(i))^2)
\biggr \} \biggr ) 
\label{a0}
\ea
Attention thus focusses on the term $- {\rm tr} \ {\rm trg} \ln {\rm
  N}(J)$ as given by Eqs.~(\ref{eq36}) to (\ref{eq43}), with the
understanding that the substitutions~(\ref{eq72}) have been made.

We recall that we deal with the case $v \ll \lambda$. Indeed, in the
calculation of localization properties, it is customary to consider
only the zeroth order contribution in $v/\lambda$ to the source term.
In the calculation of the loop corrections, we follow this usage. It
turns out that for a complete understanding of localization
properties, terms of order $v^2$ are also important. These are studied
in Section~\ref{loop1}.

Thus, the loop expansion is generated by putting $v = 0$ in ${\bf
  N}(J)$ and expanding the logarithm both in powers of $J$ and in
powers of the $\delta \sigma$'s and $\delta P$'s. We denote the terms 
generated in this way by $(p,q)$ where $p$ denotes the combined power
of the $\delta \sigma$'s and $\delta P$'s, and where $q$ denotes the
power of $J$. We obviously need to consider only the cases where $q =
0, 1, 2$. The expansion is simplified because of the form of
Eqs.~(\ref{eq72}) and (\ref{a5}): The factors $T$ and $T^{-1}$ cancel
everywhere except in the source terms $J$, and the expansion proceeds
effectively in powers of $g_{\rm d} \delta P$, $g_{\rm d} \delta
\sigma$, and $g_{\rm d} T^{-1} J T$. Here the $g_{\rm d}$'s are
defined in Eqs.~(\ref{a5}).

We evaluate first the lowest--order terms and, later, turn to the
general form of the loop expansion. We consider the terms $(p,q)$ with
$p \leq 4$ and $q \leq 2$. The term $(0,0)$ vanishes identically. As
remarked above, the term $(1,0)$ approximately cancels against the
linear terms stemming from the quadratic expressions in the exponent
of the generating function. The terms $(p,1)$ vanish for $p < n$.
Upon expansion of the exponential and calculation of the Gaussian
integrals, all terms which are odd in the $\delta \sigma$'s and
$\delta P$'s vanish. This leaves us with the following combinations of
brackets (up to fourth order in total).
\ba
&& (0,2) \ ; \ (2,2) \ ; \ (4,2) \ ; \ (1,2)(3,0) \ ; \ (2,2)(2,0) \ ;
(0,2)(2,0) \ ; \nonumber \\
&&\qquad \qquad \ (0,2)(2,0)(2,0) \ ; \ (0,2)(4,0) \ . 
\label{a7}
\ea
In writing the terms $(2,2)(2,0)$ and $(0,2)(2,0)(2,0)$ we imply that
we also expand the second--order terms $(2,0)$ originating from the
expansion of the logarithm into a Taylor series. This procedure
differs from the standard loop expansion where such second--order
terms are added to the ones which stem from the quadratic terms in the
exponent of Eq.~(\ref{eq72}). We do so for technical reasons. We shall
see later that this step does not affect our conclusions. The term
$(0,2)$ was calculated in Section~\ref{non}. We focus attention on the
remaining terms.

It is useful to work out the Gaussian integrals first (prior to taking
the traces in Hilbert space). The Gaussian integrals can be performed
using Wick contraction~\cite{itz}. We use that
\ba
\overline{ {\rm trg} ( \delta \sigma A ) {\rm trg} ( \delta \sigma B )
    } &=& \frac{1}{N} {\rm trg} (A B) \ , \nonumber \\
\overline{ {\rm trg} ( \delta \sigma A \delta \sigma B ) } &=&
  \frac{1}{N} {\rm trg} ( A ) {\rm trg} ( B ) \ .
\label{a8c}
\ea
These rules apply for arbitrary graded matrices $A$ and $B$ and can
easily be checked by direct calculation. The symbol $\delta \sigma$
stands for $\delta P$ or for any of the $\delta \sigma$'s, and $N$ is
any of the large factors multiplying the quadratic terms in
Eq.~(\ref{a0}). Needless to say, the pair of $\delta \sigma$'s must
carry equal indices $(a,b; \tau, i)$.

Using these rules it is easy to see that all terms listed
in~(\ref{a7}) vanish identically except for the term $(4,2)$. This is
because after applying the rules~(\ref{a8c}) until all $\delta P$'s
and $\delta \sigma$'s have disappeared, we are left with a product of
graded traces each involving powers of the propagators $g^{(aa)}_{\rm
  d}$ or $g^{(a(a+1))}_{\rm d}$. But each of these two propagators
has the form $F_{\mu \nu} \delta_{\alpha \beta} + G_{\mu \nu} L_{\alpha
  \beta}$. Here, $F$ and $G$ are matrices in Hilbert space, and the
indices $\alpha$ and $\beta$ refer to the graded space. Thus, the
graded traces of the $g$'s vanish, and the same statement holds for
arbitrary powers of these propagators. For later use, we keep track of
the result,
\be
{\rm trg} [ ( g^{(aa)}_{\rm d} )^k ] = 0 = {\rm trg} [ (
g^{(a(a+1))}_{\rm d} )^k ]
\label{a7a}
\ee 
which applies for all non--negative integers $k$.

The term $(4,2)$ can be worked out explicitly. A completely analogous
calculation was done in Ref.~\cite{ben} and is not repeated here. One
finds that the term $(4,2)$ does not vanish for $l \rightarrow \infty$
but is of the same order of magnitude as the term $(0,2)$ worked out
in Section~\ref{non}. This finding is consistent with the general
result of Ref.~\cite{ben} which states that if the interaction has
rank $k$ and the number of Fermions is $m$ then the terms in the loop
expansion do not vanish if $2k \leq m$. In the present case we have $k
= 1$ and $m = 2$. We conclude that our result Eq.~(\ref{eq77}) for the
localization length may be modified by the loop expansion. This is why
we now examine terms of arbitrary order in the loop expansion. We do
so in order to anwer the question: Does the loop expansion contribute
terms which depend on the distance between the box $(a(a+1)$ and the
box $((a+n)(a+n+1))$, i.e., between the end points of the Green's
function in Eq.~(\ref{a1})? If the answer is yes, the behavior of the
localization length displayed in Eq.~(\ref{eq77}) will change;
otherwise, the result in Eq.~(\ref{eq77}) is not affected by the loop
expansion.

Prior to performing the Wick contractions, the terms in the loop
expansion have the form
\be
(p_1,q_1) (p_2,q_2) \times \ldots \times (p_n,q_n) \ .
\label{a8b}
\ee
Here $\sum_i q_i = 2$ while $\sum_i p_i = p \geq 2$ must be even. We
have omitted a binomial factor which is due to the Taylor expansions
of both, the logarithm and the exponential. We refer to expressions
like~(\ref{a8b}) as to ``terms'', and to the individual factors
$(p_i,q_i)$ as to ``brackets''. Each bracket with $q_i = 0$ comprises
a sum over all boxes. Moreover, each bracket contains a trace over
Hilbert space. Therefore, each term in the sum over boxes defines one
or several closed loops in the sequence of boxes. For instance, the
bracket $(3,0)$ allows for the realization
\ba
&&\sum_a \ {\rm tr} \ {\rm trg} \ [ \ g^{((a-1)a)}_{\rm d} \
\sum_{\tau} \ \lambda \ C^{(1 \tau)}(-1,0;l) \ \delta \Sigma^{((a-1)a;
  \tau)} \ g^{(aa)}_{\rm d} \nonumber \\
&&\times \ \sum_{\tau'} \ \lambda \ C^{(1 \tau')}(0,1;l) \ \delta
\Sigma^{(a(a+1); \tau')} \ g^{(a(a+1))}_{\rm d} \nonumber \\
&&\times \ \sum_{\tau''} \ \lambda \ C^{(1 \tau'')}(-1,1;r) \ [ \delta
\Sigma^{((a-1)(a+1); \tau'')} ]^{\dagger} \ ] \ .
\label{a9b}
\ea
Here and in the sequel, we use the notation introduced in
Eqs.~(\ref{eq46}). In the expression~(\ref{a9b}), the loop consists of
the sequence $((a-1)a) \rightarrow (aa) \rightarrow (a(a+1))
\rightarrow ((a-1)a)$. Another possible realization of the bracket
$(3,0)$ is given by
\be
\sum_a \ {\rm tr} \ {\rm trg} \ [ \ g^{(aa)}_{\rm d} \delta P^{(aa)} \
g^{(aa)}_{\rm d} \delta P^{(aa)} \ g^{(aa)}_{\rm d} \delta P^{(aa)} \
] \ .
\label{a10b}
\ee
Here the loop runs entirely within the box $(aa)$. In general, we
speak of a realization of a term if, together with the sequence of
brackets~(\ref{a8b}), also the sequence of $\delta P$'s and $\delta
\sigma$'s is fixed.

The $\delta P$'s and $\delta \sigma$'s come in two classes. The first
class comprises the forms 
\be
\delta P^{(aa)} \ , \ \delta P^{(a(a+1))} \ , \ \sum_{\tau} C^{(1
  \tau)}(0) \ \delta \sigma^{(aa, \tau)} \ , \ {\rm and} \ \sum_{\tau}
C^{(1 \tau)}(1) \ \delta \sigma^{(a(a+1), \tau)} \ .
\ee
These forms connect states within a box to states within the same
box. We refer to these forms jointly by the symbol $\delta D$ where
$D$ stands for ``diagonal''. The second class comprises the forms
\ba
&&\sum_{\tau} C^{(1 \tau)}(-1,0;l) \ \delta \Sigma^{((a-1)a, \tau)} \ ,
\ \sum_{\tau} C^{(1 \tau)}(0,1;l) \  \delta \Sigma^{(a(a+1), \tau)} \
, \ {\rm and} \nonumber \\
&&\qquad \sum_{\tau} C^{(1 \tau)}(-1,1;l) \ \delta
\Sigma^{((a-1)(a+1), \tau)}
\ea
which connect different boxes and move upward in the sequence of
boxes, and the forms
\ba
&&\sum_{\tau} C^{(1 \tau)}(-1,0;r) \ [ \delta \Sigma^{((a-1)a, \tau)}
\ ]^{\dagger}\ , \ \sum_{\tau} C^{(1 \tau)}(0,1;r) \ [ \ \delta
\Sigma^{(a(a+1), \tau)} \ ]^{\dagger} \ , \ {\rm and} \nonumber \\
&&\qquad \sum_{\tau} C^{(1 \tau)}(-1,1;r) \ [ \ \delta
\Sigma^{((a-1)(a+1), \tau)} \ ]^{\dagger}
\ea
which move downward in the sequence of boxes. We refer to the first
(second) group jointly by the symbol $\delta \Sigma$ ($\delta
\Sigma^{\dagger}$, respectively). Within each realization of a term
of the loop expansion, the number of $\delta D$'s must be even, and to
each $\delta \Sigma$ there must correspond a $\delta \Sigma^{\dagger}$
of the same type.

For each term in the loop expansion, we first study all the
realizations which do not carry any $\delta \Sigma$'s and $\delta
\Sigma^{\dagger}$'s. For these realizations, the brackets $(p_i,1)$
vanish because the source term $J$ connects the boxes $(a(a+1))$ and
$((a+n)(a+n+1))$ with $n \neq 0$, and there is no way of constructing a
closed loop from $J$ and the $\delta D$'s. We conclude that only terms
of the form $(p_1,0) \times \ldots \times (p_{n-1},0) (p_n,2)$ possess
the realizations here considered. The brackets $(p_i,0)$ define closed
loops within the same box, with a final summation over all boxes. The
bracket $(p_n,2)$ consists of a loop with $p_n - k$ steps within the
box $(a(a+1))$, followed by the step $(a(a+1)) \rightarrow
((a+n)(a+n+1))$ induced by $J$, followed by $k$ steps within the box
$((a+n)(a+n+1))$, followed by the step $((a+n)(a+n+1)) \rightarrow
(a(a+1))$ induced by the second factor $J$. Here, $k = 0,1, \ldots,
p_n$. In every bracket $(p_i,0)$ with $i = 1, \ldots, (n-1)$ we split
the sum over boxes into two parts. The first part comprises the sum
over the boxes $(a(a+1))$ and $((a+n)(a+n+1))$ and is denoted by
$(p_i,0)_1$. The sum over all remaining boxes is denoted by
$(p_i,0)_2$. The product $(p_1,0) \times \ldots \times (p_{n-1},0)$ is
accordingly given by a contribution of the form $(p_1,0)_1 \times
\ldots \times (p_{n-1},0)_1$ and another contribution where at least
one of the brackets $(p_i,0)$ carries the index $2$. The Gaussian
average over the second contribution multiplied by $(p_n,2)$ vanishes.
Indeed, since each $\delta D$ is multiplied from the left by a factor
$g$, application of the contraction rules~(\ref{a8c}) leaves us, after
all Wick contractions have been performed, with a product of graded
traces of the form~(\ref{a7a}) and, thus, yields zero. We are left
with the contribution $(p_1,0)_1 \times \ldots \times (p_{n-1},0)_1
(p_n,2)$. This contribution will, in general, differ from zero and be
of the same order in $N_{a(a+1)}$ as the bracket $(0,2)$. The term
$(4,0)$ furnishes an example. The chain of terms $(4,2); (4,2)(2,0);
(6,2); (6,2)(2,0); (8,2); \ldots$ shows that terms of arbitrary order
in $p$ do contribute. Needless to say, we are not able to work out the
contributions from all these terms analytically. Performing the traces
in Hilbert space is one of the stumbling blocks. We observe, however,
that these contributions have the same form for any value of $n$. This
is because the propagators $g$, the coefficients $C^{(1 \tau)}$ and
the weight factors $N$ in the quadratic forms~(\ref{a0}) are the same
for all boxes $((a+n)(a+n+1))$. Therefore, the realizations which do
not carry any factors $\delta \Sigma$ and $\delta \Sigma^{\dagger}$
will renormalize the value of the average two--point function but will
not affect the dependence of this function on $n$. In other words, the
ratio~(\ref{eq77}) of localization lengths remains unaffected by such
realizations of the loop expansion.

We turn to the realizations which carry at least one factor $\delta
\Sigma$ and $\delta \Sigma^{\dagger}$ each  and show that for $n \geq
3$, the contributions from such realizations vanish. For any such
realization, we Wick--contract pairs of $\delta \Sigma$'s and $\delta
\Sigma^{\dagger}$'s until only a single such pair is left, leaving the
$\delta D$'s untouched. In general, there are many different ways of
performing these Wick contractions. Therefore, every realization gives
rise to many different expressions with a single remaining pair
$\delta \Sigma, \delta \Sigma^{\dagger}$ each. Each such expression
can be written in terms of the brackets introduced above. Because of
the contraction rules~(\ref{a8c}), the bracket structure is quite
different from that of the original term~(\ref{a8b}). Our claim is
proved if we show that every sequence of brackets involving two $J$'s
and a single pair $\delta \Sigma, \delta \Sigma^{\dagger}$ vanishes
upon Wick contraction.

The pair $\delta \Sigma, \delta \Sigma^{\dagger}$ connects two
different boxes which we denote by $(a,b)$ and $(a',b')$. For $n \geq
3$, at least one of these two boxes must differ from both, the box
$(a(a+1))$ and the box $((a+n)(a+n+1))$. Without loss of generality we
assume this to be the box $(a,b)$. For $n \geq 3$ both $\delta \Sigma$
and $\delta \Sigma^{\dagger}$ must occur within the same bracket.
Otherwise, it is not possible to form a closed loop. This bracket
contains either the string of factors
\be
\ldots \delta \Sigma \sum_{\mu} |ab \mu \rangle \langle ab \mu | (
g^{(ab)}_{\rm d} \delta D^{(ab)} )^k g^{(ab)}_{\rm d} \sum_{\nu} |ab
\nu \rangle \langle ab \nu | \delta \Sigma^{\dagger} \ldots
\label{a11a}
\ee
with $k$ zero or positive integer, or the Hermitean adjoint of the
form~(\ref{a11a}). According to the contraction rules~(\ref{a8c}),
Wick contraction of $\delta \Sigma$ and $\delta \Sigma^{\dagger}$
yields the factor ${\rm trg} [ ( g^{(ab)}_{\rm d} \delta D^{(ab)} )^k
g^{(ab)}_{\rm d} ] $. The remaining brackets contain the $J$'s
quadratically, an arbitrary number of $\delta D$'s, but no further
$\delta \Sigma$'s or $\delta \Sigma^{\dagger}$'s. Wick--contracting
all the $\delta D^{(ab)}$'s and application of the result~(\ref{a7a})
yields zero since the box $(ab)$ differs from both $(a(a+1))$ and
$((a+n)(a+n+1)$. This establishes our claim.

The difference between the realizations which do not carry any factors
$\delta \Sigma$ and $\delta \Sigma^{\dagger}$ and those which do, lies
in the fact that the former possess one variant (the combination
$(p_1,0)_1 \times \ldots \times (p_{n-1},0)_1 \ (p_n,2)$ where the
$\delta D$'s all carry the box indices $(a(a+1))$ or $((a+n)(a+n+1))$
while for the latter, the occurrence of at least one pair $(\delta
\Sigma, \delta \Sigma^{\dagger})$ necessarily causes the occurrence of
$\delta D$'s carrying a box index different from both $(a(a+1))$ and
$((a+n)(a+n+1))$. With the help of Eq.~(\ref{a7a}), the Gaussian
average over the latter yields zero while in the former case, the last
graded trace involves the $J$'s and, therefore, does not vanish.

In conclusion, we have shown that the result~(\ref{eq77}) for the
ratio of localization lengths remains valid to all orders in the loop
expansion.

\section{Loop Expansion: Terms of Order $v^2$}
\label{loop1}

At first sight, the result of Section~\ref{loop} is somewhat puzzling.
Indeed, it was pointed out at the end of Section~\ref{non} that the
result for the ratio of the localization lengths, Eq.~(\ref{eq77}),
cannot be exact. This provided the motivation for the investigation of
the loop expansion in Section~\ref{loop} to zeroth order in $v$. It
turned out, however, that Eq.~(\ref{eq77}) remains valid without
modification to all orders in that loop expansion. This points to the
need to investigate other loop corrections (which are definitely
needed to yield a physically acceptable result). Such corrections must
obviously be of order $v^2$ and will now be studied. They come in two
classes. There are corrections of order $v^2$ to the source terms, and
there are other corrections which are independent of the source terms.
It is easy to see that corrections in the first class cannot correct
the result Eq.~(\ref{eq77}) in a physically acceptable way. Indeed,
the transport coefficient in Eq.~(\ref{eq76a}) is proportional to
$\Delta_0 \ \Delta_1$. This factor must be corrected to become
independent of the saddle--point solution. We note that this factor
applies to all pairs of boxes $[(aa),(a(a+1)]$ and $[((a-1)a),(aa)]$,
for $a = 1, \ldots, K$, and so must, therefore, the correction we are
looking for. But by definition of the hopping matrix elements, a term
of order $v^2$ which modifies the source term can affect only the pair
of boxes immediately adjacent to the end--point boxes that occur in
Eq.~(\ref{a1}). In keeping with common practice we do not, therefore,
consider corrections in the first class any further.

It turns out that terms in the second class do provide the required
corrections. To study such terms, we return to ${\rm tr} \ {\rm trg}
\ln {\bf N}(J)$ as defined in Eqs.~(\ref{eq36}) to (\ref{eq43}) with
the substitutions~(\ref{eq72}). To calculate terms of order $v^2$
which are independent of $J$, it suffices to consider block--diagonal
submatrices $M^{(aa)}$ and $M^{((a-1)a)}$ of ${\bf N}(0)$. Here
$M^{(aa)}$ contain the boxes $(aa)$ and $(a(a+1))$ in both 
rows and columns,
\ba
\langle a a \mu | M^{(aa)} | a a \nu \rangle &=& (E - \lambda
\sigma^{(aa)}_{\rm sp} - \lambda T^{(aa)} \delta P^{(aa)}
(T^{(aa)})^{-1} ) \delta_{\mu \nu} \nonumber \\
&&- w_{\mu \nu} \nonumber \\
&&- \sum_{\tau} \lambda C^{(1 \tau)}_{\mu \nu}(0)
T^{(aa)} \delta \sigma^{(aa; \tau)} (T^{(aa)})^{-1} \ , \nonumber \\
\langle a a \mu | M^{(aa)} | a (a+1) \nu \rangle &=& -v \delta_{\mu
  \nu} - (i/2) \sum_{\tau} \lambda C^{(1 \tau)}_{\mu \nu}(-1,0;l)
\nonumber \\
&&\times T^{(aa)} \delta \Sigma^{(a(a+1); \tau)} (T^{(a(a+1))})^{-1} \
, \nonumber \\
\langle a (a+1) \mu | M^{(aa)} | a a \nu \rangle &=& -v \delta_{\mu
  \nu} + (i/2) \sum_{\tau} \lambda C^{(1 \tau)}_{\mu \nu}(-1,0;r)
\nonumber \\
&&\times T^{(aa)} [\delta \Sigma^{(a(a+1); \tau)}]^{\dagger}
(T^{(a(a+1))})^{-1} \ , \nonumber \\
\langle a (a+1) \mu | M^{(aa)} | a (a+1) \nu \rangle &=& (E - \lambda
\sigma^{(a(a+1))}_{\rm sp} \nonumber \\
&&- \lambda T^{(a(a+1))} \delta P^{(a(a+1))} (T^{(a(a+1))})^{-1} )
\delta_{\mu \nu} \nonumber \\
&& - \lambda \sum_{\tau} C^{(1 \tau)}_{\mu \nu}(1) T^{(a(a+1))} \delta
\sigma^{(a(a+1); \tau)} \nonumber \\
&&\qquad \times (T^{(a(a+1))})^{-1} \ ,
\label{l2}
\ea
while $M^{((a-1)a)}$ contains the boxes $((a-1)a)$ and $(aa)$ in both
rows and columns and is not displayed explicitly. For the terms of
second order in $v$, we have
\be
{\rm tr} \ {\rm trg} \ln {\bf N}(0) \approx \sum_a {\rm tr} \ {\rm
  trg} \ [ \ln \ M^{(aa)} + \ln M^{((a-1)a)} \ ] \ .  
\label{l1}
\ee
To justify Eq.~(\ref{l1}), we observe that contributions which are
correct in second order in $v$ can be generated from the matrix
${\bf N}(0)$ by dropping in that matrix all elements $v$ but two. The
set of matrices generated in this way, when inserted into the
expression ${\rm tr} \ {\rm trg} \ln$ and expanded into powers of the
$v$'s, $\delta P$'s and $\delta \sigma$'s, produces the same terms of
second order in $v$ as ${\rm tr} \ {\rm trg} \ln \ {\bf N}(0)$ itself.
In those matrices where the two remaining $v$'s do not appear in
mirror positions with respect to the main diagonal (as they do in
Eqs.~(\ref{l2})), the terms in the power series will not carry all
$\delta \Sigma$'s and $[ \delta \Sigma ]^{\dagger}$'s in pairs of the
same type and will, therefore, vanish upon Wick contraction. Those
matrices which do have the two remaining $v$'s appear in mirror
positions with respect to the main diagonal will have much larger
size than shown in Eqs.~(\ref{l2}), containing in addition other boxes
connected with the ones shown in Eqs.~(\ref{l2})) by $\delta \Sigma$'s
and $[ \delta \Sigma ]^{\dagger}$'s. The arguments presented towards
the end of Section~\ref{loop} (see the discussion around
expression~(\ref{a11a})) apply likewise in the present case, however,
and show that contributions from such additional $\delta \Sigma$'s and
$[ \delta \Sigma ]^{\dagger}$'s vanish. We are left with matrices
which are block--diagonal, the diagonal block shown explicitly in
Eqs.~(\ref{l2}) being disconnected from the remaining one(s). When we
evaluate the expression ${\rm tr} \ {\rm trg} \ln$, the two diagonal
blocks contribute additively, and only $M^{(aa)}$ produces
contributions of second order in $v$. We see that Eq.~(\ref{l1}) holds
except for terms which vanish upon Wick contraction and in the sense
that the terms of second order generated by expanding either side,
agree.

We focus attention on $\ln \ M^{(aa)}$; the terms $\ln M^{((a-1)a)}$
are treated analogously. Expanding $\ln \ M^{(aa)}$ in powers of $v$
and the $\delta P$'s and $\delta \sigma$'s, keeping only the terms of
second order in $v$ (but terms of arbitrary order in the $\delta P$'s
and $\delta \sigma$'s), we obtain a series of terms which individually
have the form
\be
v^2 \ {\rm tr} \ {\rm trg} [ A^{(aa)} (T^{(aa)})^{-1} (T^{(a(a+1))})
A^{(a(a+1))} (T^{(a(a+1))})^{-1} (T^{(aa)}) ] \ . 
\label{l3}
\ee
The matrix $A^{(aa)}$ has the form (we drop the Hilbert space indices
$(\mu,\nu)$) 
\ba
A^{(aa)} &=& ( g^{(aa)}_{\rm d} ( \lambda \delta P^{(aa)} +
\sum_{\tau} C^{(1 \tau)}(0) \delta \sigma^{(aa; \tau)} )^{k_1}
g^{(aa)}_{\rm d} \nonumber \\
&& \times \sum_{\tau} \lambda C^{(1 \tau)}(-1,0;l) \delta
\Sigma^{(a(a+1); \tau)} \nonumber \\
&&\times ( g^{(a(a+1))}_{\rm d} ( \lambda \delta P^{(a(a+1))} +
\sum_{\tau} C^{(1 \tau)}(1) \delta \sigma^{(a(a+1); \tau)} )^{k_2}
g^{(a(a+1))} \nonumber \\
&& \times \sum_{\tau} \lambda C^{(1 \tau)}(-1,0;r) [ \delta
\Sigma^{(a(a+1); \tau)} ]^{\dagger} \ldots \ ,
\label{l4}
\ea
and correspondingly for $A^{(a(a+1))}$. The exponents $k_1, k_2,
\ldots$ are positive integer or zero. In both $A^{(aa)}$ and
$A^{(a(a+1))}$, $\delta \Sigma$'s and $\delta \Sigma^{\dagger}$'s with
equal indices must always occur in pairs.

We expand the exponent of $\overline{Z}$ in powers of the
terms~(\ref{l3}), keep only terms of second order in $v$, perform the 
Wick contractions, and re--exponentiate the result. This procedure is
correct to order $v^2$. It is easy to see that after Wick contraction,
each of the terms~(\ref{l3}) either vanishes or is proportional to
\be
(v/\lambda)^2 \ {\rm trg} ( T^{(aa)} L (T^{(aa)})^{-1} T^{(a(a+1))} L
(T^{(a(a+1))})^{-1} ) \ ,
\label{l5}
\ee
and that the dimensionless constant of proportionality does not depend
on the box labels $(aa)$ and $(a(a+1))$. We conclude that our
procedure does generate the expected renormalization of the transport
coefficient $(v/\lambda)^2 \ \Delta_0 \ \Delta_1$ appearing in
Eq.~(\ref{eq75}). It remains to determine the magnitude of the
renormalization.

Unfortunately, this cannot be done using the loop expansion. Even the
analytical calculation of individual terms beyond the ones of lowest
order (let alone that of the entire series) is a prohibitive task. The
difficulty lies in performing the traces in Hilbert space. Therefore,
we resort to a different approach. We use a perturbative expansion
which, in a similar context, was developed in Ref.~\cite{agassi} and
which is akin to diagrammatic impurity perturbation theory. This
approach cannot give us an explicit expression either for the
renormalized transport coefficient. It does allow us, however, to
understand the physical origin and meaning of the renormalization
terms.

Starting point is the observation that in Eqs.~(\ref{l2}), only two
boxes play a role. Instead of the original Hamiltonian $H$ in
Eq.~(\ref{eq00}), we, therefore, consider the projection $PHP$ of $H$
onto the part of Hilbert space spanned by the states $| aa \mu
\rangle$ and $| a(a+1) \mu \rangle$. We consider the average
two--point function
\be
\overline{| \langle aa \mu_0 | (E^+ - PHP)^{-1} | a(a+1) \nu_0 \rangle
  |^2} \ .
\label{l6}
\ee
To calculate this expression, we first imagine using the technique
developed in the present paper. The resulting averaged generating
function would be written as an integral over the variables $T^{(aa)},
T^{(a(a+1))}, \delta P^{(aa)}, \delta P^{(a(a+1))}, \delta
\sigma^{(aa; \tau)}, \\ \delta \sigma^{(a(a+1); \tau)}, \delta
\Sigma^{(a(a+1); \tau)}$ and $[\delta \Sigma^{(a(a+1);
  \tau)}]^{\dagger}$. Except for a source term which reflects the 
choice~(\ref{l6}) of observable and is irrelevant for what follows,
the exponent of the generating function would contain the term $- {\rm
  tr} \ {\rm trg} \ \ln M^{(aa)}$ with $M^{(aa)}$ defined in
Eq.~(\ref{l2}). This fact then connects our renormalization problem
to the two--point function defined in~(\ref{l6}). Put differently,
determining our renormalized transport coefficient is tantamount to
determining the transport coefficient connecting the boxes $(aa)$ and
$(a(a+1))$ for the average two--point function~(\ref{l6}).

An alternative way of doing the calculation consists in expanding
$(E^+ - PHP)^{-1}$ in Eq.~(\ref{l6}) in powers of the stochastic
variables $h^{(a)}_{ij}$ in Eq.~(\ref{eq1}). Using the fact
that the $h^{(a)}_{ij}$'s are Gaussian distributed random variables
with zero mean value, the terms in the resulting series can be averaged
using Wick contraction. The resulting contraction patterns come in two
classes: (i) those where all contraction lines connect partners within
the same box and (ii) those where at least one contraction line
connects a partner in box $(aa)$ with a partner in box $(a(a+1))$.
Eq.~(\ref{eq8c}) shows that the second class is not empty. Iteration of
the Pastur equation, Eq.~(\ref{eq51}), and comparison with our
perturbation expansion shows that the saddle--point solution comprises
all contraction patterns in class (i) with non--intersecting
contraction lines. This is a well--known result. In the terminology of
Section~\ref{loop}, the sum of the remaining contraction patterns
corresponds to the sum of terms in the loop expansion. Realizations of
terms in the loop expansion which are of order zero in $\delta \Sigma$
and $[\delta \Sigma]^{\dagger}$ correspond to the remaining contraction
patterns in class (i). The rest corresponds to class (ii). It was
shown in a different context~\cite{ben} that the remaining
contraction patterns in class (i) are needed to change the average
level density as given by the saddle--point condition, into the true
level density. It is natural to assume that the same statement applies
in the present context. Contraction patterns in class (ii) are not
encountered frquently. They arise here because the interactions in
boxes $(aa)$ and $(a(a+1))$ are correlated, and so must be the level
densities $\rho_0(E)$ and $\rho_1(E)$.

As a result of these heuristic arguments, we are led to the conclusion
that the loop expansion renormalizes the transport coefficient. The
saddle--point version $T_{\rm sp}$ given in Eq.~(\ref{eq76a}) is
changed into
\be
T = 2 \pi \ v^2 \ \langle \rho_{0}(E) \rho_{1}(E) \rangle \ .
\label{l8}
\ee
The brackets indicate the average over the product of correlated
densities. The ratio~(\ref{eq77}) of localization lengths changes
accordingly into 
\be
\frac{\zeta(w \neq 0)}{\zeta(w = 0)} = \frac{\langle \rho_{0}(E)
\rho_{1}(E) \rangle}{\langle (\rho_{1}(E))^2 \rangle } \ . 
\label{l7}
\ee
We now take the limit in which the longitudinal thickness $d$ of the
slices tends to zero while their number $K$ goes to infinity. This
yields
\be
\frac{\zeta(w \neq 0)}{\zeta(w = 0)} = {\rm lim}_{d \rightarrow
  0} \biggl ( \frac{\langle \rho_{0}(E) \rho_{1}(E) \rangle}{\langle
(\rho_{1}(E))^2 \rangle } \biggr ) \ . 
\label{l9}
\ee
We expect that the ratio on the right--hand side of Eq.~(\ref{l9})
changes very smoothly with $d$ and that a good numerical
approximation can be obtained for finite slide thickness.

While we have reached the end of what seems possible by analytical
means, the arguments presented above suggest a numerical approach to
the problem which is feasible and which can be used to determine both
$T$ and the ratio $\zeta(w \neq 0) / \zeta(w = 0)$ for any given
two--body interaction $w$. For each of the two boxes $(aa)$ and
$(a(a+1))$, we normalize the level density so that $\int {\rm d}E
\rho(E) = 1$. We use $\rho(E) = (i/(2 \pi)) (1/N) {\rm tr} [ G^+(E) -
G^-(E) ]$ where $N$ stands for the dimension of Hilbert space in
either box. Then we can write $\langle \rho_0(E) \rho_1(E) \rangle$
in the form
\ba
&&\langle \rho_0(E) \rho_1(E) \rangle = - \frac{1}{4 \pi N_{aa}
N_{a(a+1)}} \nonumber \\
&&\qquad \qquad \times \langle {\rm tr}[G^+_{aa}(E) - G^-_{aa}(E)]
{\rm tr} [G^+_{a(a+1)}(E) - G^-_{a(a+1)}(E)] \rangle \ .
\label{ll}
\ea
The $w$--dependence rests in $[G^+_{aa}(E) - G^-_{aa}(E)]$. The
numerical calculation would draw the one--body Hamiltonians for the
two particles in the slices $a$ and $(a+1)$ at random, determine the
Green's functions $G^+_{aa}(E)$ and $G^+_{a(a+1)}(E)$ by diagonalization
of the resulting Hamiltonians in the two boxes, form the expression
${\rm tr}[G^+_{aa}(E) - G^-_{aa}(E)] \ {\rm tr}[G^+_{a(a+1)}(E) -
G^-_{a(a+1)}(E)]$, and repeat the procedure many times so as to generate
a meaningful ensemble average. In contrast to a numerical simulation of
the full problem involving a large number of boxes, this approach is
quite feasible. The simplification which we have achieved by analytical
means consists in the restriction of the calculation to two boxes.

\section{Discussion and Summary}
\label{disc}

We have studied the localization properties of two interacting
electrons in a disorder potential. Both electrons move within a quasi
one--dimensional wire. This wire is thought of as being divided into
$K$ slices, with the surfaces separating neighboring slices
perpendicular to the axis of the wire. The electrons interact if in
the same slice. Hopping matrix elements allow each electron to move
from any slice to either of the two neighboring slices. To simplify
the problem, we have admitted only those states in Hilbert space where
both electrons either occupy the same or two neighboring slices. We
have argued that this simplification should be physically irrelevant:
The influence of the two--body interaction on localization properties
should not depend qualitatively on the omission of states in Hilbert
space where the two electrons are two or more slices apart. The limit
in which the longitudinal thickness of each slice tends to zero and
the number $K$ of slices tends to infinity is considered but turns out
not to affect our results in an essential way.

The analytical treatment of this problem becomes possible thanks to
the eigenvalue decomposition of the second moment of the matrix
elements of the disorder Hamiltonian in Eqs.~(\ref{eq16}). This is the
novel technical feature of our approach. It allows us to use the
supersymmetry technique. We calculate the average of the generating
function, use the Hubbard-Stratonovich transformation, and determine
the saddle--point solutions in more or less standard fashion. The
result~(\ref{eq76a}) for the transport coefficient and for the 
ratio~(\ref{eq77}) for the localization lengths in the presence and in
the absence of the two--body interaction $w$ is intuitively convincing
and demonstrates the influence of $w$.

For the shape of the spectral density of two non--interacting
electrons, the saddle--point solution yields the semicircle, a result
which is manifestly not exact. This led us to consider loop
corrections to the saddle--point solution. Herein lies the second
technically novel aspect of our work. We succeeded in exhibiting
general properties of all terms in the loop expansion up to arbitrary
order. We could show that the loop corrections to the source terms do
not alter the results for the transport coefficient and, thus, for the
localization length obtained in the framework of the saddle--point
approximation. We studied further loop corrections which are
independent of the source terms. For these terms we could show to
arbitrary order in the loop expansion that each of the contributions
which are of second order in $v$ has precisely the form needed to
yield a renormalization correction to the transport coefficient. We
were not able to calculate the coefficients multiplying these terms
and, thus, the magnitude of the renormalization effect. Instead, we
used heuristic arguments to show that the renormalization correction
does correspond to physical expectations.

Our main result is embodied in Eq.~(\ref{l9}). This is the third novel
aspect of our work. We have shown analytically that the two--body
interaction does affect the localization length. It does so because it
alters the level density for those states in Hilbert space where the
two electrons occupy the same slice. The result may be an increase or
decrease of the localization length, depending both on properties of
the two--body interaction and on the location of the energy $E$ in the
spectrum. A sizeable change occurs if the two--body interaction meets
the criterion~(\ref{crit}). Effectively, this puts an upper bound on
the strength of the impurity potential which can easily be checked in
practice.

We draw attention to a characteristic difference between the
one--electron and the two--electron problem. In the former case, the
dimensionless transport coefficient is given in the saddle--point
approximation by $2 \pi v^2 (\Delta_1(E))^2$. The energy $E$ is that
of the electron. The result is exact because loop corrections vanish
in the limit $N \to \infty$. In the case of two electrons without
interaction, the dimensionless transport coefficient has, in the
saddle--point approximation, the form $2 \pi v^2 (\Delta_1(E))^2$.
The apparent similarity of both expressions is deceiving because now
$E$ is the total energy of the two electrons. Moreover, the result
is not exact but modified by contributions from the loop expansion
which changes the form of the level density into a convolution of
two semicircles. We might, of course, specify the energy of each of
the two electrons. But this is not a meaningful thing to do if we
wish to compare the localization lengths without and with interaction.
In the latter case, the total energy is the only constant of motion.

The supersymmetry method does not apply to one--dimensional problems.
Therefore, we cannot compare our result with what has been found
numerically in one--dimensional systems. It is possible, however, to
compare our work with the result Eq.~(\ref{0}) which is not
restricted to one dimension. This expression does not contain the
hopping matrix element $v$. Thus, it differs from our
expression~(\ref{l8}) in which we retain the structure typical for the
Thouless block scaling argument with hopping between boxes but modify
the level densities. Moreover, the arguments presented in the previous
paragraph show that it is not straightforwardly possible to compare
the localization length of the two--electron problem with that of the
one--electron problem. We note, however, that we predict a change of
localization length which depends upon all moments ${\rm tr}(w),
{\rm tr}(w^2), \ldots$ of the two--body interaction while the result
Eq.~(\ref{0}) involves only $U^2$ and is, thus, independent of the
sign of the two--body interaction.

To the best of our knowledge, this is the first time that a complete
analytical treatment of the combined effects of disorder and
interaction has been possible. We believe to have given a complete
analysis of the influence of the two--body interaction on localization
properties of two electrons in a quasi one--dimensional disorder
potential. This statement is subject to the proviso that we have
worked in a reduced Hilbert space.

\section{Appendix 1: Eigenvectors and Eigenvalues of the matrices
  ${\cal A}^{(i)}$}

The matrices ${\cal A}^{(-1)}$ and ${\cal A}^{(-1,0)}$ were treated in
Section~\ref{mom}. The remaining matrices are treated correspondingly.
Therefore, we simply list eigenvectors and left-- and right--hand
eigenvalues.

\ba
&&i = (-1): \ \Lambda^{(0)}(-1) = 2l ; \ C^{(0)}_{\mu \nu}(-1;r) =
C^{(0)}_{\mu \nu}(-1;l) \propto \delta_{\mu \nu} \ ; \nonumber \\
&&\qquad \qquad \qquad \Lambda^{(1)}(-1) = 2l-1 ; \ C^{(1 \tau)}_{\mu
  \nu}(-1;l) = (C^{(1 \tau)}_{\mu \nu}(-1;r))^{\dagger} \nonumber \\
&&\qquad \qquad \qquad \propto \langle (a-1) a \mu |
\alpha^{\dagger}_{c m} \alpha_{c m'} | (a-1) a \nu \rangle \ {\rm with}
\ c = (a-1),a \ ;
\nonumber \\ 
&&i = (0): \ \Lambda^{(0)}(0) = 2(l-1) ; \ C^{(0)}_{\mu \nu}(0;r) =
C^{(0)}_{\mu \nu}(0;l) \propto \delta_{\mu \nu} \nonumber \\
&&\qquad \qquad \qquad \Lambda^{(1)}(0) = l-2 ; \ C^{(1 \tau)}_{\mu
  \nu}(0;l) = (C^{(1 \tau)}_{\mu \nu}(0;r))^{\dagger} \nonumber \\
&&\qquad \qquad \qquad \propto \langle a a \mu |
\alpha^{\dagger}_{a m} \alpha_{a m'} | a a \nu \rangle \ ; \nonumber \\
&&i = (+1): \ \Lambda^{(0)}(+1) = 2l ; \ C^{(0)}_{\mu \nu}(+1;r) =
C^{(0)}_{\mu \nu}(+1;l) \propto \delta_{\mu \nu} \ ; \nonumber \\
&&\qquad \qquad \qquad \Lambda^{(1)}(+1) = 2l-1 ; \ C^{(1 \tau)}_{\mu
  \nu}(+1;l) = (C^{(1 \tau)}_{\mu \nu}(+1;r))^{\dagger} \nonumber \\
&&\qquad \qquad \qquad \propto \langle a (a+1) \mu |
\alpha^{\dagger}_{c m} \alpha_{c m'} | a (a+1) \nu \rangle \ {\rm with}
\ c = a,(a+1) \ ; \nonumber \\
&&i = (-1,0): \ \Lambda^{(1)}(-1,0) = l-1 ; \nonumber \\
&&\qquad \qquad \qquad C^{(1 \tau)}_{\mu \nu}(-1,0;l) \propto \langle
a a \mu | \alpha^{\dagger}_{a m} \alpha_{(a-1) m'} | (a-1) a \nu
\rangle \ ; \nonumber \\
&&\qquad \qquad \qquad C^{(1 \tau)}_{\mu \nu}(-1,0;r) \propto \langle
(a-1) a \mu | \alpha^{\dagger}_{(a-1) m} \alpha_{a m'} | a a \nu
\rangle \ ; \nonumber \\
&&i = (0,-1): \ \Lambda^{(1)}(0,-1) = l-1 ; \nonumber \\
&&\qquad \qquad \qquad C^{(1 \tau)}_{\mu \nu}(0,-1;l) \propto \langle
(a-1) a \mu | \alpha^{\dagger}_{(a-1) m} \alpha_{a m'} | a a \nu
\rangle \ ; \nonumber \\
&&\qquad \qquad \qquad C^{(1 \tau)}_{\mu \nu}(0,-1;r) \propto \langle
a a \mu | \alpha^{\dagger}_{a m} \alpha_{(a-1) m'} | (a-1) a \nu
\rangle \ ; \nonumber \\
&&i = (0,+1): \ \Lambda^{(1)}(0,+1) = l-1 ; \nonumber \\
&&\qquad \qquad \qquad C^{(1 \tau)}_{\mu \nu}(0,+1;l) \propto \langle
a (a+1) \mu | \alpha^{\dagger}_{(a+1) m} \alpha_{a m'} | a a \nu
\rangle \ ; \nonumber \\
&&\qquad \qquad \qquad C^{(1 \tau)}_{\mu \nu}(0,+1;r) \propto \langle
a a \mu | \alpha^{\dagger}_{a m} \alpha_{(a+1) m'} | a (a+1) \nu
\rangle \ ; \nonumber \\
&&i = (+1,0): \ \Lambda^{(1)}(+1,0) = l-1 ; \nonumber \\
&&\qquad \qquad \qquad C^{(1 \tau)}_{\mu \nu}(+1,0;l) \propto \langle
a a \mu | \alpha^{\dagger}_{a m} \alpha_{(a+1) m'} | a (a+1) \nu
\rangle \ ; \nonumber \\
&&\qquad \qquad \qquad C^{(1 \tau)}_{\mu \nu}(+1,0;r) \propto \langle
a (a+1) \mu | \alpha^{\dagger}_{(a+1) m} \alpha_{a m'} | a a \nu
\rangle \ ; \nonumber \\
&&i = (-1,+1): \ \Lambda^{(1)}(-1,+1) = l ; \nonumber \\
&&\qquad \qquad \qquad C^{(1 \tau)}_{\mu \nu}(-1,+1;l) \nonumber \\
&&\qquad \qquad \qquad \qquad \propto \langle a (a+1) \mu |
\alpha^{\dagger}_{(a+1) m} \alpha_{(a-1) m'} | (a-1) a \nu \rangle \ ;
\nonumber \\ 
&&\qquad \qquad \qquad C^{(1 \tau)}_{\mu \nu}(-1,+1;r) \nonumber \\
&&\qquad \qquad \qquad \qquad \propto \langle (a-1) a \mu |
\alpha^{\dagger}_{(a-1) m} \alpha_{(a+1) m'} | a (a+1) \nu \rangle \ ;
\nonumber \\
&&i = (+1,-1): \ \Lambda^{(1)}(+1,-1) = l ; \nonumber \\
&&\qquad \qquad \qquad C^{(1 \tau)}_{\mu \nu}(+1,-1;l) \nonumber \\
&&\qquad \qquad \qquad \qquad \propto \langle (a-1) a \mu |
\alpha^{\dagger}_{(a-1) m} \alpha_{(a+1) m'} | a (a+1) \nu \rangle \ ;
\nonumber \\ 
&&\qquad \qquad \qquad C^{(1 \tau)}_{\mu \nu}(+1,-1;r) \nonumber \\
&&\qquad \qquad \qquad \qquad \propto \langle a (a+1) \mu |
\alpha^{\dagger}_{(a+1) m} \alpha_{(a-1) m'} | (a-1) a \nu \rangle \ .
\label{a100}
\ea
All other eigenvalues are zero. In the first three cases, special
attention must be paid to the cases where $m = m'$, see
Section~\ref{mom}. For $i = -1, 0 +1$ the normalization factors of the
Kronecker delta's are $l^2, l^2/2, l^2$, respectively. We take $l \gg
1$. Then, we have
\ba
\Lambda^{(0)}(i) &=& 2l \ {\rm for} \ i = -1, 0, +1; \nonumber \\
\Lambda^{(1)}(i) &=& l \ {\rm all} \ i \ {\rm except} \ i = \pm 1 \
{\rm where} \ \Lambda^{(1)}(i) = 2l \ .
\label{a2}
\ea

\section{Appendix 2: The second moment of expression~(\ref{eq17})}

Because of the fact that only neighboring slices are admitted in the
Hilbert space defined in Section~\ref{mod}, the expression~(\ref{eq17})
reduces to
\ba
&&- \frac{i}{2} \sum_{a=1}^{K-1} \sum_{\mu \nu} \Psi^{*}_{a (a+1) \mu}
L^{1/2} \langle a (a+1) \mu | H_0 | a (a+1) \nu \rangle L^{1/2}
\Psi_{a (a+1) \nu} \nonumber \\
&&\qquad - \frac{i}{2} \sum_{a=1}^K \sum_{\mu \nu} \Psi^{*}_{a a \mu}
L^{1/2} \langle a a \mu | H_0 | a a \nu \rangle L^{1/2} \Psi_{a a \nu}
\ .
\label{eq18}
\ea
After averaging over the ensemble $\{H_0\}$, the generating functional
contains in the exponent a term given by $1/2$ times the mean value of
the square of the expression~(\ref{eq18}). We use the notation
introduced in Section~\ref{mom} and in Appendix 1 and obtain for
this term the following contributions. There are two ``diagonal''
terms given by
\ba
&&- \frac{\lambda^2}{8l} \sum_{a=1}^{K-1} \sum_{\mu \nu \rho \sigma}
\Psi^{*}_{a (a+1) \mu} L \Psi_{a (a+1) \sigma} \Psi^{*}_{a (a+1) \rho}
L \Psi_{a (a+1) \nu} {\cal A}^{(1)}_{\mu \nu \rho \sigma}  \ , \nonumber
\\
&&- \frac{\lambda^2}{8l} \sum_{a=1}^K \sum_{\mu \nu \rho \sigma}
\Psi^{*}_{a a \mu} L \Psi_{a a \sigma} \Psi^{*}_{a a \rho} L \Psi_{a a
\nu} {\cal A}^{(0)}_{\mu \nu \rho \sigma} \ .
\label{eq19}
\ea
A diagonal term containing ${\cal A}^{(-1)}$ does not appear and would
be redundant. There are six ``non--diagonal'' terms given by
\ba
&&- \frac{\lambda^2}{8l} \sum_{a=2}^{K} \sum_{\mu \nu \rho \sigma}
\Psi^{*}_{a a \mu} L \Psi_{a a \sigma} \Psi^{*}_{a (a-1) \rho} L
\Psi_{a (a-1) \nu} {\cal A}^{(0,-1)}_{\mu \nu \rho \sigma}  \ ,
\nonumber \\
&&- \frac{\lambda^2}{8l} \sum_{a=2}^{K} \sum_{\mu \nu \rho \sigma}
\Psi^{*}_{(a-1) a \mu} L \Psi_{(a-1) a \sigma} \Psi^{*}_{a a \rho} L
\Psi_{a a \nu} {\cal A}^{(-1,0)}_{\mu \nu \rho \sigma}  \ , \nonumber
\\
&&- \frac{\lambda^2}{8l} \sum_{a=2}^{K-1} \sum_{\mu \nu \rho \sigma}
\Psi^{*}_{(a-1) a \mu} L \Psi_{(a-1) a \sigma} \Psi^{*}_{a (a+1) \rho}
L \Psi_{a (a+1) \nu} {\cal A}^{(-1,1)}_{\mu \nu \rho \sigma} \ ,
\nonumber \\
&&- \frac{\lambda^2}{8l} \sum_{a=2}^{K-1} \sum_{\mu \nu \rho \sigma}
\Psi^{*}_{a (a+1) \mu} L \Psi_{a (a+1) \sigma} \Psi^{*}_{(a-1) a \rho}
L \Psi_{(a-1) a \nu} {\cal A}^{(1,-1)}_{\mu \nu \rho \sigma}  \ ,
\nonumber \\
&&- \frac{\lambda^2}{8l} \sum_{a=1}^{K-1} \sum_{\mu \nu \rho \
sigma} \Psi^{*}_{a a \mu} L \Psi_{a a \sigma} \Psi^{*}_{a (a+1) \rho}
L \Psi_{a (a+1) \nu} {\cal A}^{(0,1)}_{\mu \nu \rho \sigma}  \ ,
\nonumber \\
&&- \frac{\lambda^2}{8l} \sum_{a=1}^{K-1} \sum_{\mu \nu \rho \sigma}
\Psi^{*}_{a (a+1) \mu} L \Psi_{a (a+1) \sigma} \Psi^{*}_{a a \rho} L
\Psi_{a a \nu} {\cal A}^{(1,0)}_{\mu \nu \rho \sigma} \ .
\label{eq20}
\ea
For the sum of the terms in expressions~(\ref{eq19}) and (\ref{eq20}),
we write somewhat symbolically
\be
- \frac{\lambda^2}{8l} \sum_a \sum_{i \neq (-1)} \ \sum_{\mu \nu \rho
  \sigma} \sum_{\alpha \beta} \Psi^{*}_{\mu \alpha} L_{\alpha \alpha}
\Psi_{\sigma \alpha} \Psi^{*}_{\rho \beta} L_{\beta \beta} \Psi_{\nu
  \beta} {\cal }{\cal A}^{(i)}_{\mu \nu \rho \sigma} \ . 
\label{eq21}
\ee
We have omitted the label indices $a, (a \pm 1)$ on the $\Psi$'s.
These, however, are implied by the factors ${\cal A}^{(i)}$. We have
likewise omitted the limits on the summation over $a$. We have instead
indicated the summation over the graded indices $\alpha$ and
$\beta$. We now use the decomposition in Eq.~(\ref{eq16}) for the
${\cal A}^{(i)}$'s and obtain several contributions. The first one is
due to the terms with $s = 0$ and given by
\be
\frac{1}{2 l^2} \sum_{a=1}^K {\rm trg} [ ( A^{(a a)}(0))^2] +
\frac{1}{4 l^2} \sum_{a=1}^{K-1} {\rm trg} [ ( A^{(a (a+1))}(1))^2]
\ . 
\label{eq22}
\ee
The graded matrices $A^{(a a)}_{\alpha \beta}$ and $A^{(a
  (a+1))}_{\alpha \beta}$ are defined by
\ba
A^{(a a)}_{\alpha \beta} &=& i \lambda \sum_{\mu} L^{1/2}_{\alpha
  \alpha} \Psi_{a a \mu \alpha} \Psi^{*}_{a a \mu \beta} L^{1/2}_{\beta
\beta} \ , \nonumber \\
A^{(a (a+1)}_{\alpha \beta} &=& i \lambda \sum_{\mu} L^{1/2}_{\alpha
  \alpha} \Psi_{a (a+1) \mu \alpha} \Psi^{*}_{a (a+1) \mu \beta}
L^{1/2}_{\beta \beta} \ .
\label{eq23}
\ea
The second one is due to the terms with $s = 1$ and $i = 0, i = 1$. We
find
\be
\frac{1}{4 l^2} \sum_{a=1}^K \sum_{\tau} {\rm trg} [ ( A^{(a
  a; \tau)})^2] + \frac{1}{4 l^2} \sum_{a=1}^{K-1} \sum_{\tau} {\rm
  trg} [ ( A^{(a (a+1); \tau)})^2] \ . 
\label{eq22a}
\ee
The graded matrices $A^{(a a; \tau)}_{\alpha \beta}(0)$ and $A^{(a
(a+1); \tau)}_{\alpha \beta}(1)$ are defined by
\ba
A^{(a a; \tau)}_{\alpha \beta} &=& i \lambda \sum_{\mu \nu}
L^{1/2}_{\alpha \alpha} \Psi_{a a \nu \alpha} \Psi^{*}_{a a \mu \beta}
L^{1/2}_{\beta \beta} C^{(1 \tau)}_{\mu \nu}(0) \ , \nonumber
\\ A^{(a (a+1); \tau)}_{\alpha \beta} &=& i \lambda \sum_{\mu \nu}
L^{1/2}_{\alpha \alpha} \Psi_{a (a+1) \nu \alpha} \Psi^{*}_{a (a+1)
\mu \beta} L^{1/2}_{\beta \beta} C^{(1 \tau)}_{\mu \nu}(1) \ .
\label{eq23a}
\ea
The terms with $s = 1$, $i = (-1,0)$ and $i = (0,-1)$ are given by
\ba
&&\frac{1}{8} \sum_{a = 2}^K N(-1,0)^{-1} \sum_{\tau} {\rm trg}
[A^{((a-1) a; \tau)}(l) A^{((a-1) a; \tau)}(r) \nonumber \\
&&\qquad + A^{((a-1) a, \tau)}(l) A^{((a-1) a; \tau)}(r)] \ . 
\label{eq24}
\ea
We have used that $N(-1,0) = N(0,-1)$. The graded matrices are defined
by
\ba
A^{((a-1) a; \tau)}_{\alpha \beta}(-1,0;l) &=& i \lambda \sum_{\mu \nu}
L^{1/2}_{\alpha \alpha} \Psi_{a a \nu \alpha} \Psi^{*}_{(a-1) a \mu
\beta} L^{1/2}_{\beta \beta} C^{1 \tau}_{\mu \nu}(-1,0;l) \ ,
\nonumber \\
A^{((a-1) a; \tau)}_{\alpha \beta}(-1,0;r) &=& i \lambda \sum_{\mu \nu}
L^{1/2}_{\alpha \alpha} \Psi_{(a-1) a \nu \alpha} \Psi^{*}_{a a \mu
\beta} L^{1/2}_{\beta \beta} C^{1 \tau}_{\mu \nu}(-1,0;r) \ ,
\nonumber \\
A^{((a-1) a; \tau)}_{\alpha \beta}(0,-1;l) &=& i \lambda \sum_{\mu \nu}
L^{1/2}_{\alpha \alpha} \Psi_{(a-1) a \nu \alpha} \Psi^{*}_{a a \mu
\beta} L^{1/2}_{\beta \beta} C^{1 \tau}_{\mu \nu}(0,-1;l) \ ,
\nonumber \\
A^{((a-1) a; \tau)}_{\alpha \beta}(0,-1;r) &=& i \lambda \sum_{\mu \nu}
L^{1/2}_{\alpha \alpha} \Psi_{a a \nu \alpha} \Psi^{*}_{(a-1) a \mu
\beta} L^{1/2}_{\beta \beta} C^{1 \tau}_{\mu \nu}(0,-1;r) \ .
\nonumber \\
\label{eq25}
\ea
The terms with $s = 1$ and $i = (0,1)$ and $i = (1,0)$ yield
correspondingly
\ba
&&\frac{1}{8} \sum_{a = 1}^{K-1} N(1,0)^{-1} \sum_{\tau} {\rm trg}
[A^{(a (a+1); \tau)}(0,1;l) A^{(a (a+1); \tau)}(0,1;r) \nonumber \\
&&\qquad + A^{(a (a+1), \tau)}(1,0;l) A^{((a-1) a; \tau)}(1,0;r)] \
,
\label{eq26}
\ea
with
\ba
A^{(a (a+1); \tau)}_{\alpha \beta}(0,1;l) &=& i \lambda \sum_{\mu \nu}
L^{1/2}_{\alpha \alpha} \Psi_{a (a+1) \nu \alpha} \Psi^{*}_{a a \mu
\beta} L^{1/2}_{\beta \beta} C^{1 \tau}_{\mu \nu}(0,1;l) \ ,
\nonumber \\
A^{(a (a+1); \tau)}_{\alpha \beta}(0,1;r) &=& i \lambda \sum_{\mu \nu}
L^{1/2}_{\alpha \alpha} \Psi_{a a \nu \alpha} \Psi^{*}_{a (a+1) \mu
\beta} L^{1/2}_{\beta \beta} C^{1 \tau}_{\mu \nu}(0,1;r) \ ,
\nonumber \\
A^{(a (a+1); \tau)}_{\alpha \beta}(1,0;l) &=& i \lambda \sum_{\mu \nu}
L^{1/2}_{\alpha \alpha} \Psi_{a a \nu \alpha} \Psi^{*}_{a (a+1) \mu
\beta} L^{1/2}_{\beta \beta} C^{1 \tau}_{\mu \nu}(1,0;l) \ ,
\nonumber \\
A^{(a (a+1); \tau)}_{\alpha \beta}(1,0;r) &=& i \lambda \sum_{\mu \nu}
L^{1/2}_{\alpha \alpha} \Psi_{a (a+1) \nu \alpha} \Psi^{*}_{a a \mu
\beta} L^{1/2}_{\beta \beta} C^{1 \tau}_{\mu \nu}(1,0;r) \ .
\nonumber \\
\label{eq27}
\ea
Finally, the terms with $s = 1$ and $i = (-1,1)$ and $i = (1,-1)$
yield
\ba
&&\frac{1}{8} \sum_{a = 2}^{K-1} N(-1,1)^{-1} \sum_{\tau} {\rm trg}
[A^{((a-1) (a+1); \tau)}(-1,1;l) A^{((a-1) (a+1); \tau)}(-1,1;r)
\nonumber \\ 
&&\qquad + A^{((a-1) (a+1), \tau)}(1,-1;l) A^{((a-1) (a+1);
  \tau)}(1,-1;r)] \ ,
\label{eq28}
\ea
with
\ba
A^{((a-1) (a+1); \tau)}_{\alpha \beta}(-1,1;l) &=& i \lambda \sum_{\mu
\nu} L^{1/2}_{\alpha \alpha} \Psi_{a (a+1) \nu \alpha} \Psi^{*}_{(a-1)
a \mu \beta} L^{1/2}_{\beta \beta} C^{1 \tau}_{\mu
  \nu}(-1,1;l) \ ,
\nonumber \\
A^{((a-1) (a+1); \tau)}_{\alpha \beta}(-1,1;r) &=& i \lambda \sum_{\mu
\nu} L^{1/2}_{\alpha \alpha} \Psi_{(a-1) a \nu \alpha} \Psi^{*}_{a (a+1)
\mu \beta} L^{1/2}_{\beta \beta} C^{1 \tau}_{\mu
  \nu}(-1,1;r) \ ,
\nonumber \\
A^{((a-1) (a+1); \tau)}_{\alpha \beta}(1,-1;l) &=& i \lambda \sum_{\mu
\nu} L^{1/2}_{\alpha \alpha} \Psi_{(a-1) a \nu \alpha} \Psi^{*}_{a (a+1)
\mu \beta} L^{1/2}_{\beta \beta} C^{1 \tau}_{\mu
  \nu}(1,-1;l) \ ,
\nonumber \\
A^{((a-1) (a+1); \tau)}_{\alpha \beta}(1,-1;r) &=& i \lambda \sum_{\mu
\nu} L^{1/2}_{\alpha \alpha} \Psi_{a (a+1) \nu \alpha} \Psi^{*}_{(a-1)
a \mu \beta} L^{1/2}_{\beta \beta} C^{1 \tau}_{\mu \nu}(1,-1;r) \ .
\nonumber \\
\label{eq29}
\ea
Eqs.~(\ref{a100}) show that $C^{(1 \tau)}_{\mu \nu} (-1,0;l) = C^{(1
  \tau)}_{\mu \nu} (0,-1;r)$ and correspondingly for the pairs $i =
(0,1), i = (1,0)$ and $i = (-1,1), i = (1,-1)$. The definitions in
Eqs.~(\ref{eq25},\ref{eq27}) and (\ref{eq29}) imply that
$A^{((a-1) a; \tau)}(-1,0;l) = A^{((a-1) a; \tau)}(1,0;r)$ and
similar relations for the remaining five pairs of $A$'s. Therefore,
the sum of the terms in expressions~(\ref{eq24},\ref{eq26},\ref{eq28})
simplifies to
\ba
&&\frac{1}{4} \sum_{a = 2}^K N(-1,0)^{-1} \sum_{\tau} {\rm trg}
[A^{((a-1) a; \tau)}(-1,0;l) A^{((a-1) a; \tau)}(-1,0;r) \nonumber \\
&&+ \frac{1}{4} \sum_{a = 2}^{K-1} N(-1,1)^{-1} \sum_{\tau} {\rm trg}
[A^{((a-1) (a+1); \tau)}(-1,1;l) A^{((a-1) (a+1);
  \tau)}(-1,1;r)\nonumber \\
&&+ \frac{1}{4} \sum_{a = 1}^{K-1} N(1,0)^{-1} \sum_{\tau} {\rm trg}
[A^{(a (a+1); \tau)}(0,1;l) A^{(a (a+1); \tau)}(0,1;r) \ .
\label{eq30}
\ea
The sum of the terms in expressions~(\ref{eq22},\ref{eq22a}) and
(\ref{eq30}) is equal to $1/2$ times the second moment of
expression~(\ref{eq17}). This is the result given in Eq.~(\ref{eq17a}).

\newpage

\begin{figure}
\center
\includegraphics[width=10cm]
{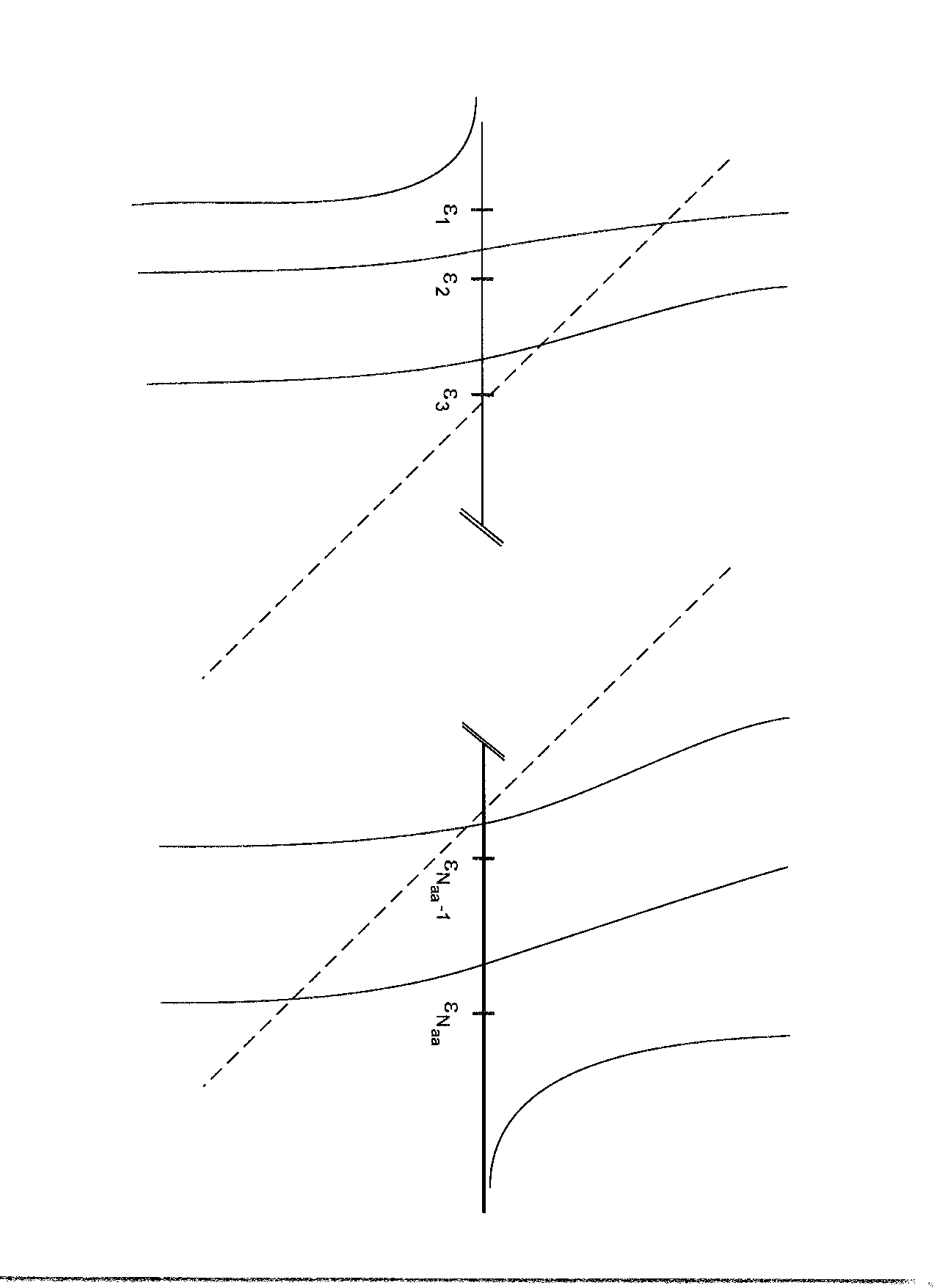}
\caption{Graphical solution (schematic) of the saddle--point
  equation, Eq.~(\ref{eq67}). Some of the eigenvalues $\varepsilon_j$
  are shown on the abscissa. The $\tau$--dependence of the right--hand
  side of Eq.~(\ref{eq67}) is indicated by the solid lines. The two
  dashed straight lines represent the left--hand side of
  Eq.~(\ref{eq67}) for two values of $z$.}
\label{fig:Figure1}
\end{figure}

\end{document}